\documentclass[journal]{IEEEtran}

\usepackage{cite}

\ifCLASSINFOpdf
   \usepackage[pdftex]{graphicx}
   \DeclareGraphicsExtensions{.pdf,.jpeg,.png}
\else
\fi

\usepackage{amsmath}

\usepackage{multirow}

\begin{document}

\title{From Fundamental First-Principle Calculations to NanoEngineering Applications: Review of the NESSIE project}

\author{James Kestyn,
        Eric Polizzi% <-this % stops a space
\thanks{Department of Electrical and Computer Engineering, University of Massachusetts, Amherst,
MA, 01003 USA e-mail: epolizzi@engin.umass.edu}% <-this % stops a space
}

\maketitle

\begin{abstract}

This paper outlines how modern first-principle calculations
can
adequately address the needs for ever higher levels of numerical accuracy and high-performance in
large-scale electronic structure simulations, and pioneer
the fundamental study of
quantum many-body effects in a large number
of emerging nanomaterials.

%This paper presents a computational process that will allow 
%fundamental first-principle calculations 
%to significantly impact innovations in nanoengineering.

%Progress in large-scale DFT and real-time TDDFT simulations will significantly impact a wide range
%of important application areas in nanotechnology.
%Once fully optimized the NESSIE simulator
%could be applied
%to investigate the fundamental electronic properties of a tremendous variety of nanostructures that
%are beyond the reach of presently employed techniques. 

%%%%%%%%%%%%%%%

\end{abstract}


% \fi
\IEEEpeerreviewmaketitle

\section{Introduction}

\IEEEPARstart{T}{he technology} for electronic devices has been on a rapidly rising trajectory since the 1960s. 
 The main factor in this development has been the ability to fabricate ever smaller 
silicon CMOS devices (`Moore's Law'), 
with today's device sizes in the nanometer range.
The ability to control electronic materials and understand their properties
has been a driving force
for technological breakthroughs.
The emergence of new nanoscale materials and
devices, whose 
operating principles rely entirely on quantum effects,
necessitates a fundamental and comprehensive
understanding of the nanoscale physics of systems. 
First principle calculations offer a unique approach to
study materials that start directly from the mathematical equations describing the physical laws and do not require any empirical parameters aside from fundamental constants.
They are known as electronic structure calculations when applied to the configuration of electrons in a molecule or solid which determine most of the physical properties of matter 
 through chemical bonding.
Fundamentals laws governing the physics have been known since the beginning of the $\rm 20^{th}$ century with the development of quantum mechanics. The difficulty, then, does not lie in formulating the problem, but actually solving it.
 Atom-by-atom large-scale first-principle calculations
 have become critical
 for supplementing the experimental investigations and 
 obtaining detailed electronic structure properties and
 reliable characterization of emerging nanomaterials.
These simulations are essential to assist the every day work of numerous
engineers and scientists and can universally impact a wide range of disciplines (engineering, physics,
chemistry, and biology) that span technological fields of computing, sensing and energy.

In spite of the enormous progress that has been made
in the last few decades, the room for improvement in first-principle calculations is still significant.
Traditional numerical and modeling techniques
are indeed largely inadequate to cope with the new generation of 
challenges encountered in large-scale nanoengineering applications 
including systems with many thousand atoms.
Atomistic simulations must adapt to leverage the current needs in scalability by capitalizing on the massively parallel
capabilities of modern high-performance computing (HPC) platforms.
Additionally, well-established public or commercial software packages were originally
intended to investigate the basic electronic structure properties of materials using ground-state
calculations. They possess only limited capabilities for performing excited-state calculations that can
efficiently model and predict quantum many-body effects in emerging nanomaterials. The ability to
capture these fundamental nanophysics effects is increasingly important
for exploring and prototyping new revolutionary functional materials in nanotechnology. There is an
urgent need in nanoengineering for new quantum-based transformative solutions that
will play a key role in future electronics including plasmonics, phononics and excitonics.
Future breakthrough could enable disruptive technologies to
compete directly with CMOS or be integrated into existing systems to increase throughput and
decrease power dissipation. To this end, new one and two-dimensional nanostructures (e.g. graphene, carbon
nanotubes, $\rm MoS_2$, layered transition metal dichalcogenides)
have been the center of large research efforts, with first-principle atomistic
simulations playing a significant role.
Plasmonic devices, that rely on collective many-body effects, have also shown
promise as high frequency analog sensors to be used in bio-medical applications and telecommunications.

\begin{figure*}[htb]
\centering
\includegraphics[width=0.9\linewidth]{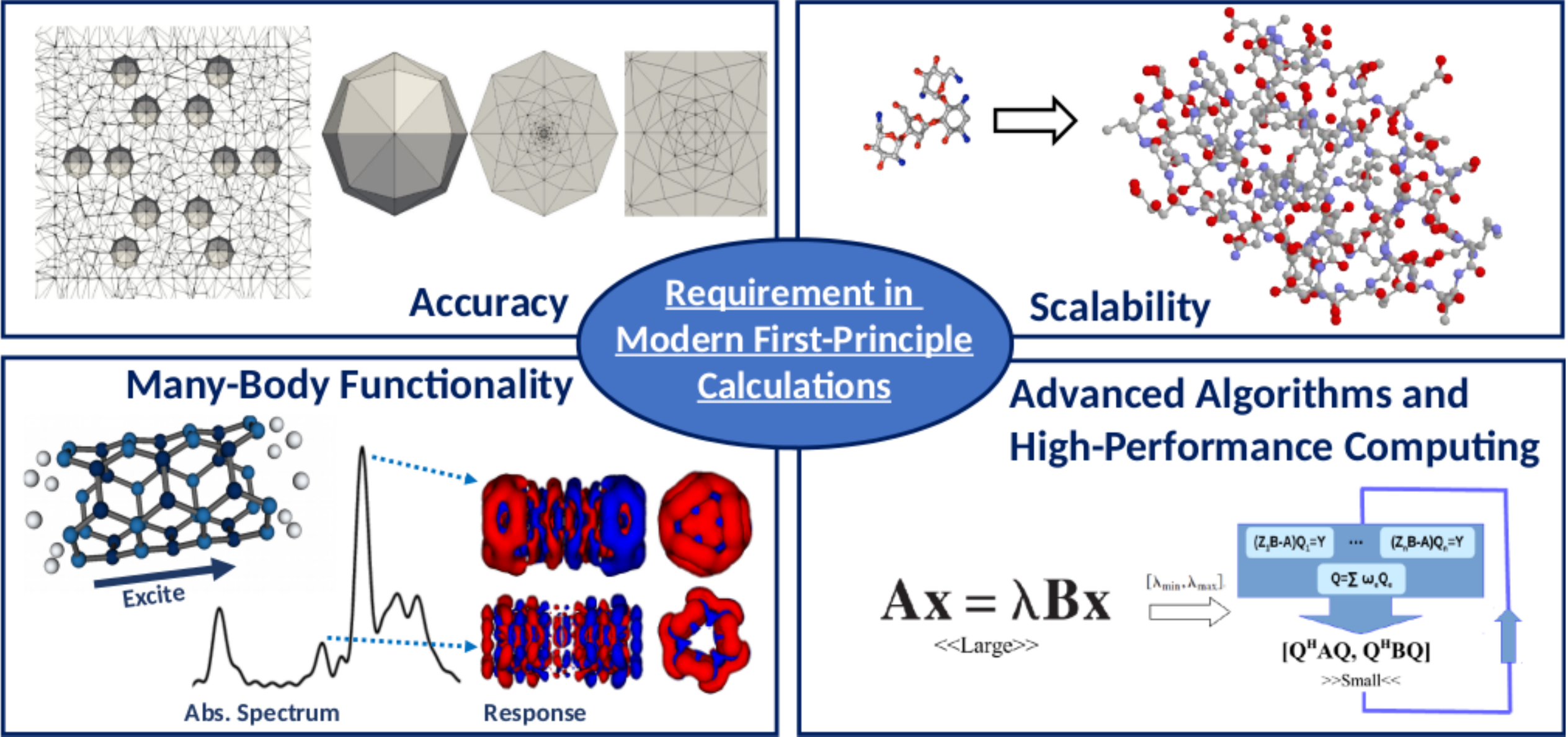}
\caption{\label{fig-modern} Requirement in modern first-principle calculations at a glance.}
\end{figure*}

This paper presents the entire computational process needed
to bring fundamental first-principle calculations 
up to the level where they can significantly impact innovations in nanoengineering.
As depicted in Figure~\ref{fig-modern}, modern first principle calculations must be capable of: (i)~achieving both accuracy and scalability;
(ii)~allowing for the study emerging many-body functionality; and (iii)~fully taking advantage of recent advanced in algorithms development
for high-performance computing (HPC). The basics of first principle modeling are first summarized in Section \ref{sec-2}.
 From ground-state to excited-state calculations, the NESSIE modeling framework is then introduced in Section  \ref{sec-3}.
NESSIE's accuracy and scalability are described in detail using various simulation results. The paper ends
by discussing plasmonics in Section \ref{sec-4} as a nanoengineering application of
large-scale first-principle calculations.

\section{First-Principle Modeling: From Physics to Algorithms}\label{sec-2}

The field of first-principle modeling can be broadly separated into three categories:
(i)~{\em physical models} to reduce the complexity of the full many-body solution while keeping much of the important physics.
 The choice of a physical model is often motivated by the objectives of the simulations;
(ii)~{\em discretization and mathematical models} that transform physical equations into the language of linear algebra;
and (iii)~{\em computing and numerical algorithms}
to solve the resulting problems.
In order to improve on current software implementation by fully capitalizing on modern HPC computing platforms, 
it is essential to revisit %not one, but
all 
the various stages of the electronic structure modeling process
which are briefly summarized in the
following.

\subsection{Physical Modeling}

%At first, the Born-Oppenheimer approximation is often used to reformulate the many-body
%problem consisting of the positions of a set of electrons and nuclei into separate electronic and nuclear
%parts.
A direct numerical treatment of a full many-body Schr\"odinger equation leads to a deceptively simple
linear eigenvalue problem, which is well known  to be intractable because of its exponential
growing dimension with the number of particles. This limitation has historically motivated the need for
lower levels of sophistication in the description of the electronic structure using a single electron picture
approximation where the size of the Hamiltonian operator ends up scaling linearly with the number
of electrons.  
First-principle electronic structure calculations 
are usually performed within the single-electron picture 
\cite{martin,Jorge} 
using either quantum chemistry (i.e. post Hartree-Fock) methods 
or, as an alternative to wave function based methods, 
Density Functional Theory (DFT) associated with the Kohn-Sham equations \cite{DFT,KS}.
Although DFT does not allow for systematic accuracy as traditional quantum chemistry techniques would, %that
%calculate the many-electron wave-function as a Slater
%determinant expansion of single particle states,
it is the method of choice when dealing with moderate sized systems containing more than a handful of atoms.
 DFT has been widely used in computational material science for decades, 
 since it provides (in principle) an exact method for calculating the ground-state
density and energy of a system of interacting electrons using a non-linear single electron Schr\"odinger-like equation associated 
with exchange-correlation (XC) functionals. In practice, the reliability of DFT depends on the numerical approximations used for the
XC terms that range from the simplest local density approximation (LDA) or the generalized gradient approximation (GGA),
to more advanced (hybrid) schemes which are still
the subject of active research efforts \cite{xcburke,SA,Vlad,kburke}.
Solutions of the DFT/Kohn-Sham problem are routinely 
used in the calculations of many ground-state properties including:
 total energy and ionization potential, crystal-atomic structure, ionic forces, vibrational frequencies, and phonon bandstructure via pertubation theory.

Although DFT cannot fundamentally provide information on excited-states and many-body properties,
the Kohn-Sham eigenvectors  are often needed by more advanced techniques: e.g.  either Green's function-based \cite{GW} (e.g. GW, Bethe-Salpeter) or time-dependent density-based 
(i.e. TDDFT\cite{TDDFTbook,alda}) approaches.  The pros and cons of these approaches are discussed in Ref.\cite{Rubio1}.
TDDFT, proposed by Runge and Gross \cite{tddft},
 continues to gain popularity as one of the most  numerically 
affordable many-body techniques capable of providing fairly accurate
results.
TDDFT has been successfully applied to calculate many physical observables of the time-dependent Hamiltonian, such as
excitation energies and complex permittivities, as well as  non-linear phenomena.  It is often used to obtain the absorption spectra of complex molecular systems.
While the design of advanced time-dependent XC functionals is still a challenging task \cite{tddftburke}, ALDA
(Adiabatic LDA) for TDDFT has been found to perform extremely well on a wide variety of systems by capturing many nanoscopic
effects (such as plasmonic effects) which, in turn, can be quantitatively compared with the experimental data.
 
TDDFT calculations can be performed in frequency or real-time domain.
The real-time TDDFT technique is a relatively recent approach introduced
by Yabana and Bertsch in \cite{YB1,YB2}, and it has  become an important focus of the TDDFT
research activities. It has notably been integrated into the software  
packages Octopus \cite{octopus,Octopus0}, NWChem \cite{nwchem}, and GPAW \cite{gpaw} for the study of molecular systems.
 In essence,  spectroscopic information can be obtained
 using the standard formalism of dipole time-response from weak short-polarized impulses
in any given direction of the system, and which requires all the occupied single electron wave functions 
to be propagated (non-linearly) in time. The imaginary part of the dipole's Fourier transform provides 
the dipole strength function. 
The absorption spectrum is then obtained along with the expected ``true many-body'' 
excited energy levels.
In contrast to the numerical models  derived from the 
TDDFT linear response theory in frequency domain \cite{TDDFTbook,Casida,tddft04,Russ-tdlda},  
the real-time TDDFT approach is better suited for achieving linear parallel scalability
and it can also address any form of non-linear responses, including ion dynamics \cite{kaxiras}.

\subsection{Mathematical Modeling and Discretization}
Although, first-principle calculations  
have provided a practical (i.e. numerical tractable) path for 
solving the electronic structure problem, they have
also introduced new numerical challenges.  
Within the single electron picture, the resulting
eigenvalue problem becomes fully non-linear since the Hamiltonian
 operator depends on all the occupied
eigenfunctions (i.e.  $H(\{\psi\})\psi=E\psi$). In practice, this
 non-linear eigenvector problem is commonly
addressed using  direct minimization schemes or self-consistent field methods (SCF) wherein 
a series of linear eigenvalue problems
(i.e.  $H\psi=E\psi$), needs to be solved iteratively until 
convergence. Computing the electron density at a
given iteration step  becomes one of the most time-consuming and challenging part
of the DFT electronic structure calculations.
Successfully reaching convergence
by performing SCF iterations is of paramount importance 
to first-principle electronic structure calculations 
software. 
%Traditional SCF mixing methods
%employ successive approximation iterates of a fixed 
%point mapping to generate the new input electron
%density at each cycle. 
%Anderson mixing, such as Newton-Broyden,
%and Pulay mixing, 
%using direct inversion of the iterative 
%subspace (DIIS) \cite{pulay,eyert2}, are common approaches.
%Such iterates, however, can be sensitive to the choice of the initial guess and they can still
%be found to converge very slowly, or not at all \cite{Yang1}.
Real-time TDDFT comes also with its own set of mathematical and numerical challenges for performing the time-propagation, 
those will be discussed further in Section \ref{sec-3}.

To perform the numerical calculations, the mathematical
 models need first to be discretized by expanding the wave
 functions over a set of basis functions. One can identify
three main discretization techniques that have been widely 
used over the past four decades by both
the quantum chemistry and the solid-state physics communities 
\cite{martin}: 
(i)~the linear combination of
atomic orbitals (LCAO) (along with the dominant use of 
Gaussian local basis sets), (ii)~the plane
wave expansion scheme, and (iii)~the real-space mesh techniques
 \cite{Beck,Batcho,White,Jim,Tsuchida,Modine,Briggs,Torsti1,Stefan,Pask,Jim2,Torsti2,allelec} 
 (also loosely called “numerical
grids”) based on finite difference method (FDM), finite element 
method (FEM), spectral element or wavelets methods.
Each of these approaches have pros and cons.
\begin{itemize}
\item Plane waves have traditionally been used within the solid-state physics community because their natural periodic nature can be easily applied to crystal structures. 
However, this can be cumbersome when dealing with finite systems where
 the computational domain must be made much larger than the molecular size to ensure interactions due to periodicity are negligible. 
Additionally, they often make use of pseudopotentials to mimic the effects of core electrons, which do not directly participate in chemical bonding and would otherwise necessitate a very large number of plane waves due to their high-frequency variations.
\item LCAO benefits from a large collection of local basis sets that 
has been improved and refined throughout the years by the 
quantum chemistry community to obtain high-level of accuracy in
simulations. However, LCAO bases may 
suffer from numerical truncation errors of finite expansions, and the solutions
cannot be universally and systematically improved towards convergence.
%often need to be 'augmented' in time-dependent simulations to capture the extended states.
\item Real-space mesh techniques provide a natural way of quantifying atomic information
 by  employing universal local mathematical approximations. 
%that can be systematically refined toward convergence. 
They can easily handle the treatment of various boundary conditions,
such as Dirichlet (for the confined directions), periodic or absorbing (for transport simulations). 
  %and they are ideally suited to be used within a real-time dependent framework (offering the same reliability
  %for capturing confined or extended states).
  Similarly to plane wave schemes, however,
  the high-level of refinement needed to capture the core electrons may be problematic.
\end{itemize}

In all cases, the level
of approximation 
in the discretization
stage, is bounded 
by the capabilities of the numerical algorithms for
solving the resulting system matrices.
In modern nanoelectronic applications, one aims at fully utilizing the power of modern
HPC architectures to tackle large-scale finite systems by exploiting parallelism at multiple levels. In this context, real-space mesh techniques offer the most significant advantages.   
They produce very sparse matrices that can take advantage 
of  recent advances made in $O(N)$ linear scaling methods
and domain decomposition techniques.

\subsection{Computing}

Much of the progress in this field is directly tied to advancements in algorithmic research
allowing larger and more complex systems to be simulated. 
Within the SCF-DFT procedure, computing the electron density by solving the linear and symmetric 
eigenvalue problem at each iteration   
becomes the major computational challenge. 
The characterization of complex systems 
and nanostructures of current
technological interests, requires the repeated computations of many tens of thousands
of eigenvectors, for eigenvalue systems that can have sizes in the tens of millions.
It is important to mention that Green's function-based  formalism %alternative %to the wave function 
%formalism
can alternatively be used for computing directly the electron
density (using efficient evaluations of the diagonal 
elements of the Green's function along a complex contour, e.g. \cite{baroni1,Lin,ZP2008b}).
However, this method gives rise to difficulties in algorithmic 
complexity (i.e. O($N^2$) for 3D systems), parallel scalability and accuracy.
In that regard, it is difficult to bypass the wave function
formalism, and progress in large-scale electronic structure 
calculations can then be tied together with 
advances in numerical algorithms for addressing the eigenvalue 
problem, in particular.

Traditional methods for solving the eigenvalue problem  (including Arnoldi, Lanczos 
methods, or other Davidson-Jacobi techniques \cite{Golub00,Template00})
and related packages \cite{freeeig}, are largely unable to cope with these challenges.
In particular, they
suffer from the orthogonalization of a very large basis when
many eigenpairs are computed. In this case, a divide-and-conquer approach that can compute 
wanted eigenpairs by parts becomes mandatory, since 'windows' or 'slices' of the
spectrum can be computed independently of one another and
orthogonalization between eigenvectors in different slices is
no longer necessary.
These issues have motivated the development of a new family of eigensolver based on contour integration
techniques  \cite{Sakurai2003,sakurai2007cirr,imakura2014block,austin2015computing} such as the FEAST eigensolver \cite{p2009,feast}. 
FEAST is an optimal accelerated
subspace iterative technique for computing interior eigenpairs making use of a rational filter to 
approximate the spectral projector \cite{tp14}. FEAST can be applied for solving both standard and
generalized forms of the Hermitian or non-Hermitian problems \cite{james2}. 
Once a given search interval is selected, FEAST's main
computational task consists of solving a set of independent linear systems along a complex contour.
Not only does the FEAST algorithm feature some remarkable and robust convergence properties \cite{tp14,zolotarev}, it can exploit natural parallelism at three different levels (L1, L2 or L3): (L1) search intervals can be treated
separately (no overlap), (L2) linear systems can be solved independently across the quadrature nodes
of the complex contour, and (L3) each complex linear system with multiple right-hand-sides can be
solved in parallel. Parallel resources can be placed at all three levels simultaneously in order to
achieve scalability and optimal use of the computing platform.

%%%%%%%%%%%%%%%%%%%%%%%%%%%%%%%%%%%%%%%%%%%%%%%%%%%%%%%%%%%%%%%%%%%%%%%%%%%%%%%%%%%%%%%%%%%%%%%%%%%%%%%%%%%%%%%%
%%%%%%%%%%%%%%%%%%%%%%%%%%%%%%%%%%%%%%%%%%%%%%%%%%%%%%%%%%%%%%%%%%%%%%%%%%%%%%%%%%%%%%%%%%%%%%%%%%%%%%%%%%%%%%%%

\section{First-Principle Calculations using NESSIE}\label{sec-3}
A first-principle simulation software must be capable of  addressing all the modern  challenges summarized
 in Figure~\ref{fig-modern}.
One of the major goal in modern first-principle calculations is to develop numerical algorithms and simulation software for electronic
structure that can scale the system size to thousands of atoms, without resorting to additional approximations beyond the DFT and TDDFT
physical models. The target computing architecture is usually comprised of thousands of processor cores and contains multiple
 hierarchical levels of parallelism.

The NESSIE project \cite{nessie} is an electronic structure code that uses a real-space finite element (FEM) discretization and domain decomposition to perform all-electron ground-state DFT and real-time excited-state TDDFT calculations. 
The code is written to take advantage of multi-level parallelisms to target systems containing many distributed-memory compute nodes. 
Custom numerical algorithms have been developed for the eigenvalue problems and linear systems representing the major linear
algebra operations within the software. 
 NESSIE's capabilities can be separated into three main categories:
 (i)~accurate large-scale full core potential DFT calculations using real space FEM discretization and domain-decomposition;
 (ii)~TDDFT real-time propagation for efficient spectroscopic calculations %from X-Ray to UV-ViS,
 allowing the study
 of many-body effects; and
 (iii)~massively parallel implementation on modern high-end computing platforms using state
 of the art parallel algorithms/solvers.

 The next sections present a step-by-step description of NESSIE's modeling framework applied to the benzene molecule as an example.

\subsection{An All-Electron HPC Framework}

\begin{figure}[htbp]
\centering
\includegraphics[width=0.9\linewidth,angle=0]{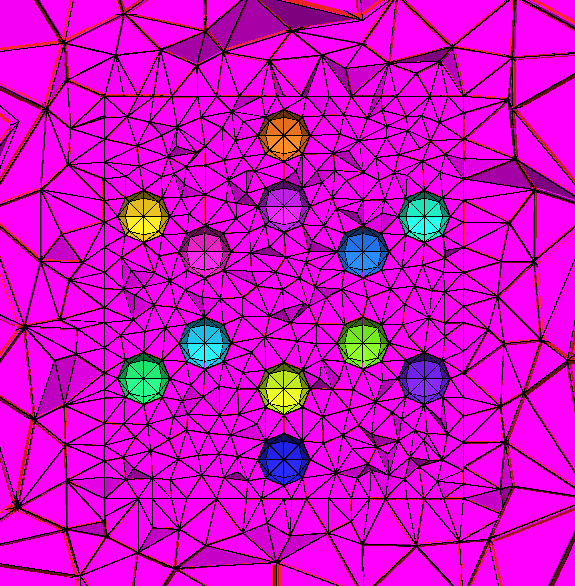} 
\begin{center}
\includegraphics[width=0.45\linewidth,angle=0]{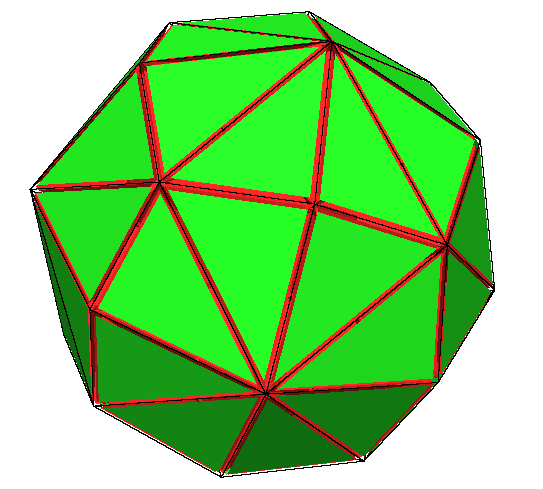}\hfill
\includegraphics[width=0.45\linewidth,angle=0]{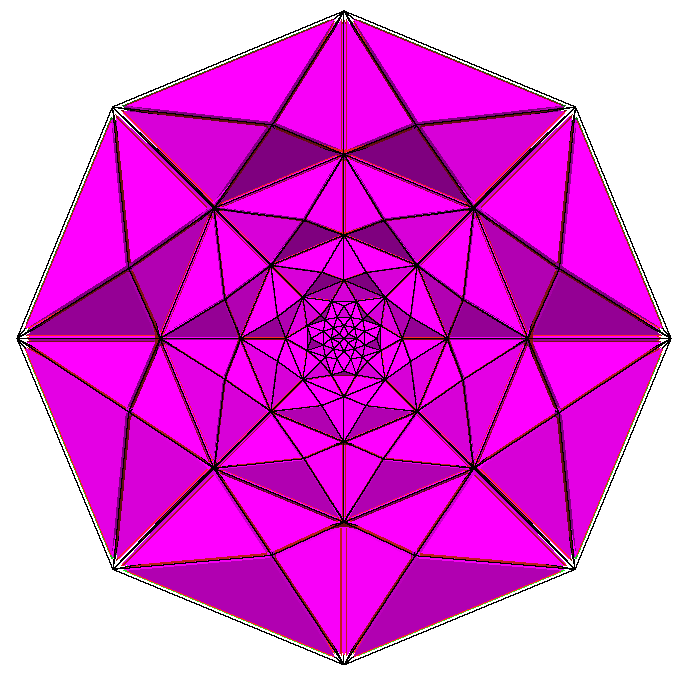}
\end{center}
\caption{ \label{fig-mesh} 
Using a muffin-tin domain-decomposition method, the whole simulation domain is separated into multiple atom-centered regions (i.e. muffins)  
and one large interstitial region. 
The top figure represents a 2D section of local FEM discretization using a coarse interstitial mesh (represented only partially here) 
connecting  all of the atoms of a Benzene molecule. The bottom figures represent a finer mesh for the atomic (muffin)  regions 
suitable to capture the highly localized core states around the nuclei.}
\end{figure}

In NESSIE, the equations
are discretized using FEM with quadratic (P2) or cubic (P3) order, along with a muffin-tin 
domain-decomposition (DD) technique.
The latter has been proposed as early as the 1930's \cite{slater}
to specifically address a multi-center atomic system.
The whole simulation domain is separated into
multiple atom-centered regions (i.e. muffins) and one large interstitial region.
 Without any loss of generality,  Figure~\ref{fig-mesh} illustrates
the essence of the muffin-tin domain decomposition using FEM and applied to 
the Benzene molecule.
The 3D finite-element muffin-tin mesh can be built in two steps: (i)~a 3D atom-centered mesh
which is  highly refined around the nucleus to capture the core states, and (ii)~a much coarser 
3D interstitial mesh that connects all the muffins (generated in NESSIE using the Tetgen software \cite{hang,tetgen}).
For the atom-centered mesh, which is common to all atoms of the same atomic number, it is convenient to use 
successive layers of polyhedra similar to the ones proposed in \cite{allelec}. 
This discretization provides both tetrahedra of good quality, and an arbitrary level of refinement i.e.
 the distance between layers can be arbitrarily refined while approaching the nucleus (this is known as a h-refinement for FEM). 

The muffin-tin decomposition can bring flexibility in the discretization step (different basis-sets can also
be used independently to describe the different regions), reduce the main computational efforts within the interstitial region alone, 
and should also guaranteed maximum linear parallel scalability performances.  
It is important to note that, independently of the type of atoms,  the outer layer of the muffin 
is consistently providing the same 
(relatively small) number of connectivity nodes $n_j$ with
 the interstitial mesh at the muffin edges (i.e. $n_j=98$, or $218$ nodes respectively using quadratic P2 or cubic P3 FEM). 
Consequently, the size of the system matrix in the interstitial region
 stays independent of  the size of the atom-centered regions, and the approach can then ideally deal 
 with full potential (all-electron).

 Once the ``Schr\"odinger'' eigenvalue problem (i.e. ${H}{\psi}=E{\psi}$) is reformulated using domain decomposition strategies,
 the resulting (and still exact) problem now takes a a non-linear form in the interstitial region 
 (i.e. ${{H_I}}(E){{\psi_I}}=E{{\psi_I}}$, since the boundary conditions at the interfaces with the muffins are energy dependent).
As originally pointed out by Slater in 1937 while introducing the muffin-tin augmented plane wave (APW) method,  
  this non-linear eigenvalue problem gives rise  to an energy dependent secular equation which cannot be handled by traditional 
  eigenvalue algorithms. 
% While introducing the augmented plane wave (APW) method using a muffin-tin domain decomposition in 1937,
% Slater originally stated \cite{slater}: 
%{\it ``Of course, we cannot solve this exactly, and we must look for methods of approximations''}.
%Indeed, this non-linear eigenvalue problem gives rise  to an energy dependent secular equation which cannot be handled by traditional 
%eigenvalue algorithms. 
Although solving such non-linear problem explicitly is not impossible \cite{harmon,sjostedt2}, 
it remains practically challenging and it is still the subject of active research efforts \cite{beyn,gavin}. 
Therefore, the mainstream approaches to all-electron  (i.e. full-potential) 
electronic structure calculations in the solid-state physics community,
have been mostly relying on approximations, such as direct 
linearization techniques, which have
been improved throughout the years  
(e.g. LAPW, LMTO, LAPW+lo,  etc.) \cite{andersen,singh2,sjostedt,madsen,bsingh}. 
Alternatively, linear eigenvalue problems can directly be obtained from pseudopotential approximation
techniques \cite{hellman,pseudo,kleinman,blochl}
 that eliminate the core states by introducing 
smooth but non-local potentials in muffin-like atom-centered regions.

\begin{figure}[htbp]
  \begin{center}
    \includegraphics[width=0.75\linewidth]{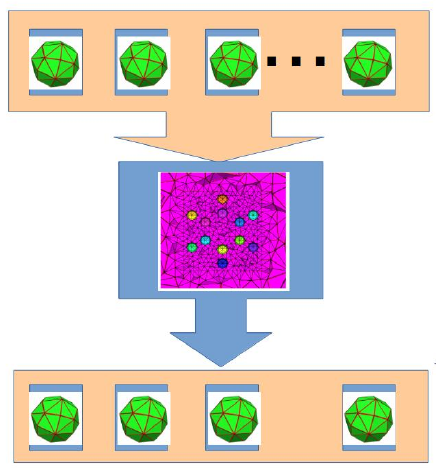}
  \end{center}
 \caption{\label{fig-muffindd} In NESSIE, the muffin-DD solver operates in three stages: 
(i)~the atoms are distributed throughout the MPI processors and factorized at a given energy pivot (i.e. shift value provided by FEAST).
The boundary conditions (i.e. self-energy) are then derived at the interfaces of the muffins.
(ii)~the resulting interstitial problem (Schur complement) is solved in parallel; and (iii)~knowing the exact solution at the muffin
interfaces, the solution within each atom is retrieved in parallel.}
\end{figure}

In NESSIE, an exact strategy has been introduced for performing all-electron
electronic structure calculations within a parallel computing environment \cite{lzp12,james}.
The approach relies on the shift-and-invert capability of  eigenvalue algorithms such as FEAST,
which leads to formulating well-defined linear systems.0
Domain decomposition methods have been well studied and are a natural framework for addressing large sparse linear systems
generated from real-space meshes. They are often associated with the use of 
distributed-memory numerical algorithms to address the data distribution using the Message Passing Interface (MPI) paradigm. 
Consequently, the solution of the FEAST's linear systems can be fully
parallelized using MPI since the muffin-tin decomposition naturally allows each muffin to be factorized 
and solve independently.
When the muffin-tin decomposition is applied to a given linear system, the resulting linear system in the interstitial domain
(a.k.a., the Schur complement) remains linear.
Figure~\ref{fig-muffindd} illustrates how the muffin-DD can be used to optimally solve the FEAST's linear systems.
The details of the muffin-DD strategy implementation have been provided in Ref. \cite{lzp12}.
 In comparison with linearization techniques discussed above, the set of `pivot energies' used to evaluate 
 the interstitial Hamiltonian system are now explicitly provided by the FEAST algorithm (i.e. they correspond
 to quadrature nodes in the complex plane) and
 they guarantee global convergence toward the correct solutions (i.e. no approximation needed).
Since the complexity of interstitial system scales linearly with the number of atoms while including non-locality only at the
interfaces with the muffins, one can also demonstrate that this all-electron framework is (paradoxically) capable
of better scalability performances than pseudopotential approaches on parallel architectures.

\begin{figure}[htbp]
 \includegraphics[width=1\linewidth]{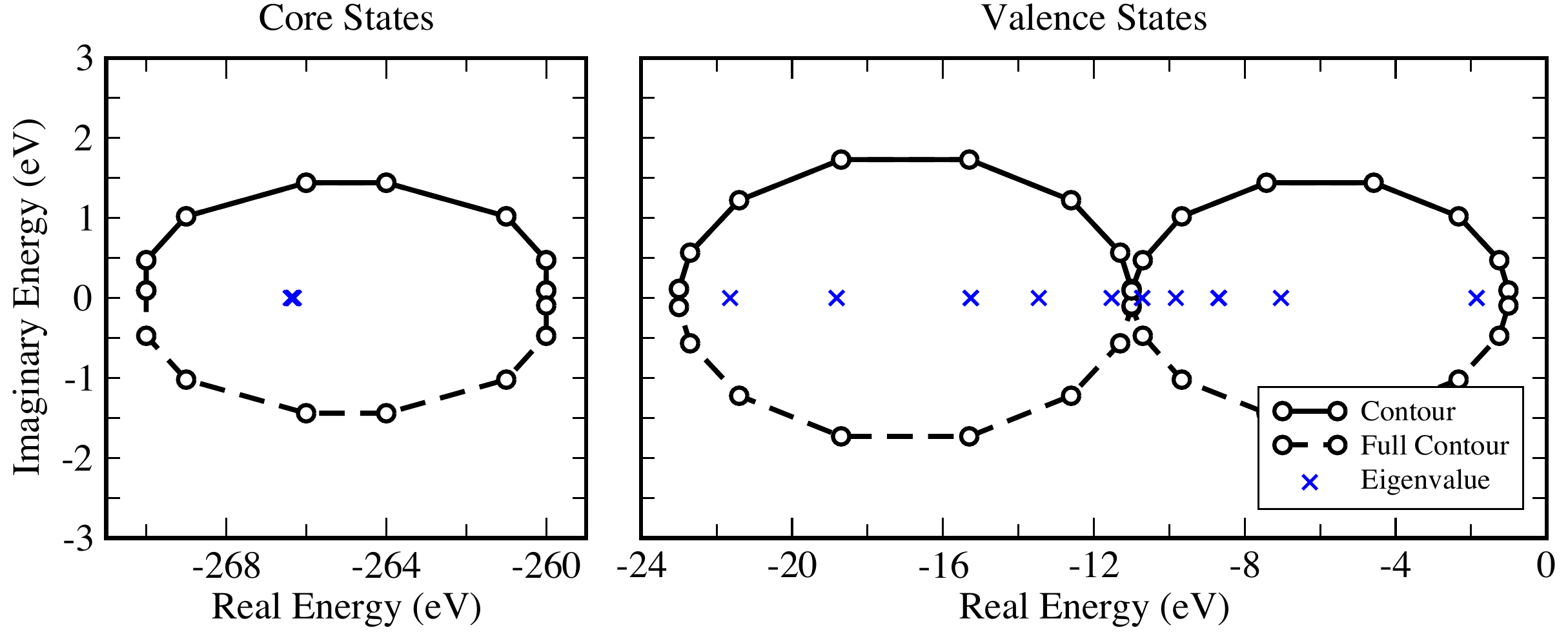}
  \caption{\em \label{fig-slicing} Splitting the full search interval into three separate contours for benzene. The lowest
   energy contour captures the core states while the other two target valence
   electrons. Each half-contour has eight quadrature nodes shown as circles. Symmetry allows
the FEAST algorithm to only perform computations  for the upper upper half of the
contour (solving eight independent linear systems per interval).}
\end{figure}

The FEAST solver has been recently undergone a significant upgrade to support the 
MPI-MPI-MPI distributed parallel programming model in v4.0 \cite{pfeast} (where the last 'MPI' refers to the linear system solves at level L3).
Furthermore, NESSIE can take advantage of the first two MPI levels of parallelism offered by FEAST 
assuming that the eigenvalue spectrum is distributed among the compute nodes. 
At the second level L2, FEAST can naturally
distribute all the linear systems associated with a given search interval (typically less than 10). 
At the first level L1,
FEAST enables `spectrum slicing' where all the intervals of interests are solved in parallel.
The use of spectrum slicing is essential to address a major bottleneck in large-scale DFT calculations
concerning the computation and storage of the DFT wave functions. The storage requirement, in particular, keeps increasing
linearly with the number of electrons in the nanostructures.
%The latter are represented by
%dense $n\times m$ matrices where the number of discretization points $n$, and the
%number of needed eigenvectors $m$, keeps increasing with the number of atoms
%in the nanostructures.
Even with simplified physical model such as DFT, it becomes particularly difficult to scale the
electronic structure problem for systems containing more than a few hundred electrons without the ability to perform spectrum slicing.
%For large scale nanostructures with more than a few hundred electrons, it becomes then
%necessary to slice the search interval for the eigenvalue problem while reducing the number of vectors stored in the wave function matrix.
This technique is illustrated for benzene in Figure~\ref{fig-slicing},
using three elliptical contours for FEAST, one that computes the core states and two others
that computes half of the valence states. In larger systems, each contour often contains hundreds of eigenvalues.

\subsection{DFT Ground-State Calculations and Scalability} 

The DFT/Kohn-Sham problem can be expressed as:
\begin{multline}
  \label{eq-scf}
{\left[-\frac{\hbar^2}{2m}\nabla^2+v_{KS}[n]({\bf r})\right]\psi_i}=E_i{\psi_i}, \\ \mbox{with} \quad {n}({\bf r})=2\sum_{i=1}^{N_e}{|\psi_i({\bf r})|^2},
\end{multline}
where the Kohn-Sham potential, $v_{KS}[n]=v_H[n]+v_{XC}[n]+v_{ext}$, is composed of 
 the Hartree potential $v_H$ solution of the Poisson equation, the XC potential, and  
 other external potential $v_{ext}$ including the ionic core potential.
 The $N_e$ lowest occupied electronic states $\{\psi_i\}_{(i=1,\dots,N_e)}$ are needed to compute the electron density
 (the factor $2$ stands for the electron spin).  
Formally, the system ({\ref{eq-scf}) forms a {\em non-linear eigenvector problem} which is commonly addressed 
using a self-consistent field method (SCF) wherein a series of linear eigenvalue problems %(i.e.  ${\bf H\Psi}=E{\bf S\Psi}$),
need to be solved iteratively until convergence.
The naive approach which consists of updating the input electron density at each SCF iteration directly from the output electron density,
results in large oscillations between SCF iterations. This approach is very unlikely to converge  as the initial guess for the density is usually far from the ground-state solution. Instead,
traditional SCF methods employ successive approximation iterates of a fixed point mapping to generate the new input electron
(a.k.a., 'mixing' techniques).

\subsubsection{Discussions on Convergence}

Using a mixing technique, two different electron densities are considered to construct the input density at the $(k+1)^{th}$ SCF iteration: the input density $\rho^k _{in}$ used to construct the Kohn Sham Hamiltonian and the output density $\rho^k _{out}$ computed from the wave functions.
With simple mixing, the input electron density for the next iteration can be computed as,
\begin{equation}
\rho^{k+1} _{in} = (1-\beta) \rho ^k _{in} + \beta \rho ^k _{out},
\end{equation}
where the parameter $\beta$ is usually chosen less than 1/2.  This, however, will converge very slowly. 
In order to increase the convergence rate, more sophisticated methods have been developed for solving this fixed-point problem. 
Newton methods cannot be used in electronic structure since it is impractical to construct the Jacobian matrix. 
Other quasi-Newton methods have been developed in the 1960s, notably by Anderson \cite{anderson1965iterative} and Broyden \cite{broyden1965class}, which do not require the Jacobian or Hessian.  
These techniques were later refined in the 1980s in the context of SCF iteration
by Pulay \cite{pulay,pulay1982improved} and have since been expanded upon \cite{kudin2002black,Yang1,walker2011anderson,gp13}. 
They are also related to Krylov methods and GMRES
\cite{saad1986gmres,eyert2,saad2010numerical}. 
For electronic structure calculations these iterative techniques are usually referred to as Direct Inversion of the
Iterative Subspace (DIIS) methods. 
The general idea is to build the input electron density as a linear combination of past densities. One can then construct 
the input mixing density,
\begin{equation}
\label{eq::rhoin}
\hat{\rho}^{k+1} _{in} = c_0 \rho^{0} _{in} + c_1 \rho^{1} _{in} + \dots + c_k \rho^{k} _{in},
\end{equation}
and the output mixing density,
\begin{equation}
\label{eq::rhoout}
\hat{\rho}^{k+1} _{out} = c_0 \rho^{0} _{out} + c_1 \rho^{1} _{out} + \dots + c_k \rho^{k} _{out},
\end{equation}
from the previous input and output densities. 
The input for the next iteration is chosen as a linear combination of the mixing densities:  
\begin{equation}
\rho^{k+1} _{in} = (1-\beta) \hat{\rho}^{k+1} _{in} + \beta \hat{\rho}^{k+1} _{out}, 
\end{equation}
where $\beta$ is again referred to as the mixing parameter.
This approach can be truncated in order to keep the density subspace size small. 
The inclusion of more densities in the mixing subspace can result in better convergence, but has diminishing returns. 
Keeping a history of ten to twenty input and output densities seems to be more than sufficient. 
Better performance can be obtained by choosing a larger value of $\beta$, but it may also result in instability.
However, improvement in convergence can be obtained by progressively increasing the $\beta$ parameter along the SCF iterations. 

The coefficients $\{c_0,...,c_k\}$ of (\ref{eq::rhoin}) and (\ref{eq::rhoout}) are the same for both the input and output mixing subspaces. 
They are computed by solving a $k \times k$ linear system with one right-hand-side, 
\begin{equation}
M_{_{k\times k}}c_{_{k\times 1}} = r_{_{k\times 1}}, 
\end{equation}
where the $i^{th}$ element of $r$ depends on the difference between the output and input densities of the current iteration $k$ and a previous iteration $i$, 
\begin{equation}
r_{i} = \int _{\Omega}  \rho ^k _{out,in} \times \left[ \rho ^k _{out,in} -  \rho ^i _{out,in}  \right] d \Omega,  
\end{equation}
with $\rho_{out,in}^p= \left( \rho ^p _{out} - \rho ^p _{in} \right)$,
and each element $M_{ij}$ of matrix $M$ takes into account the densities at iterations $i$ and $j$: 
\begin{equation}
M_{ij} = \int _{\Omega} \left[ \rho ^k _{out,in}  - \rho ^i _{out,in}  \right] \times \left[ \rho ^k _{out,in} -  \rho ^j _{out,in} \right] d \Omega. 
\end{equation}
The effect of beta mixing ratio $\beta$ and the number of density mixing subspaces kept in memory can bee seen in Figure~\ref{fig-mixing} while analyzing the convergence of benzene.
\begin{figure}[h!]
\centering
 \includegraphics[width=0.33\linewidth]{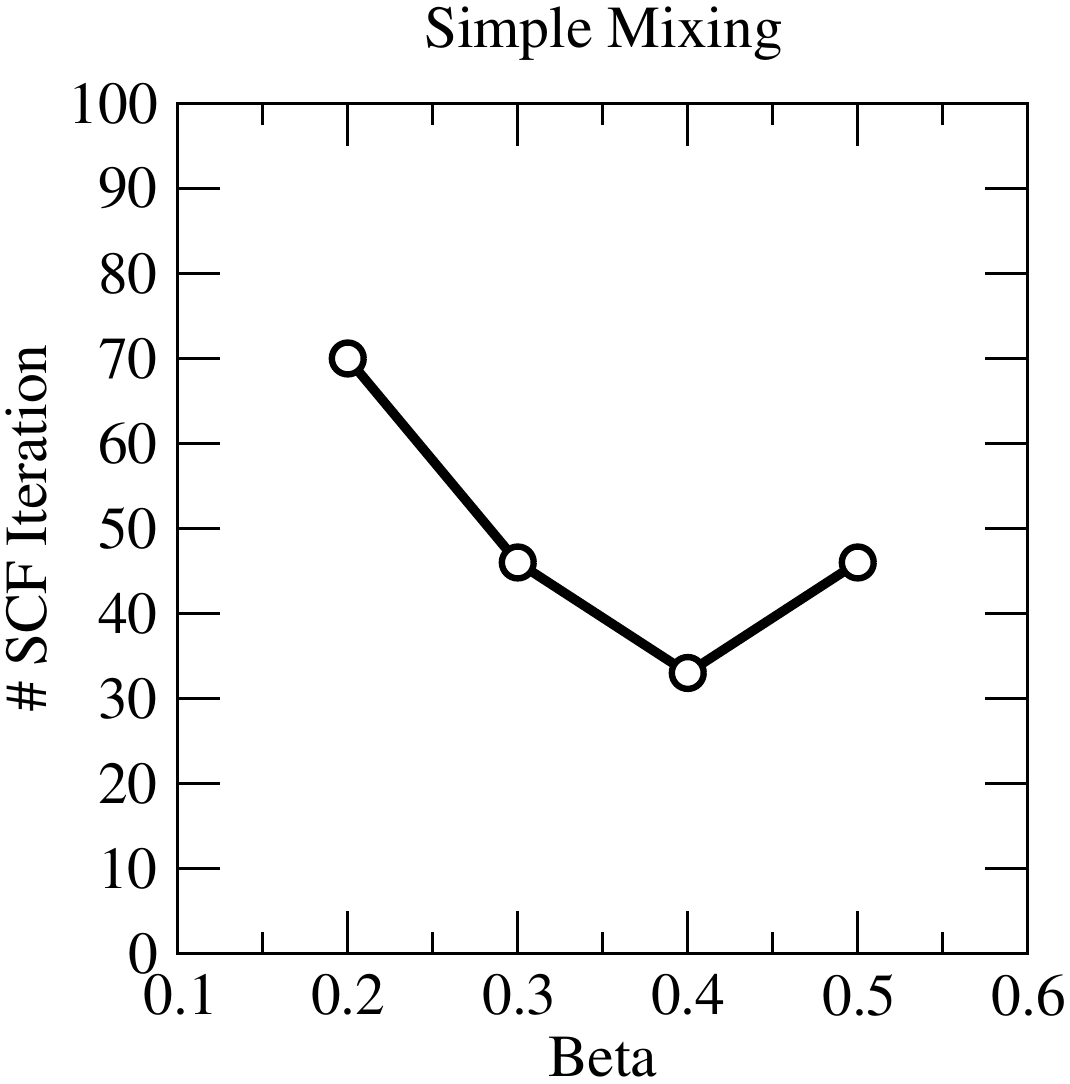}
 \includegraphics[width=0.31\linewidth]{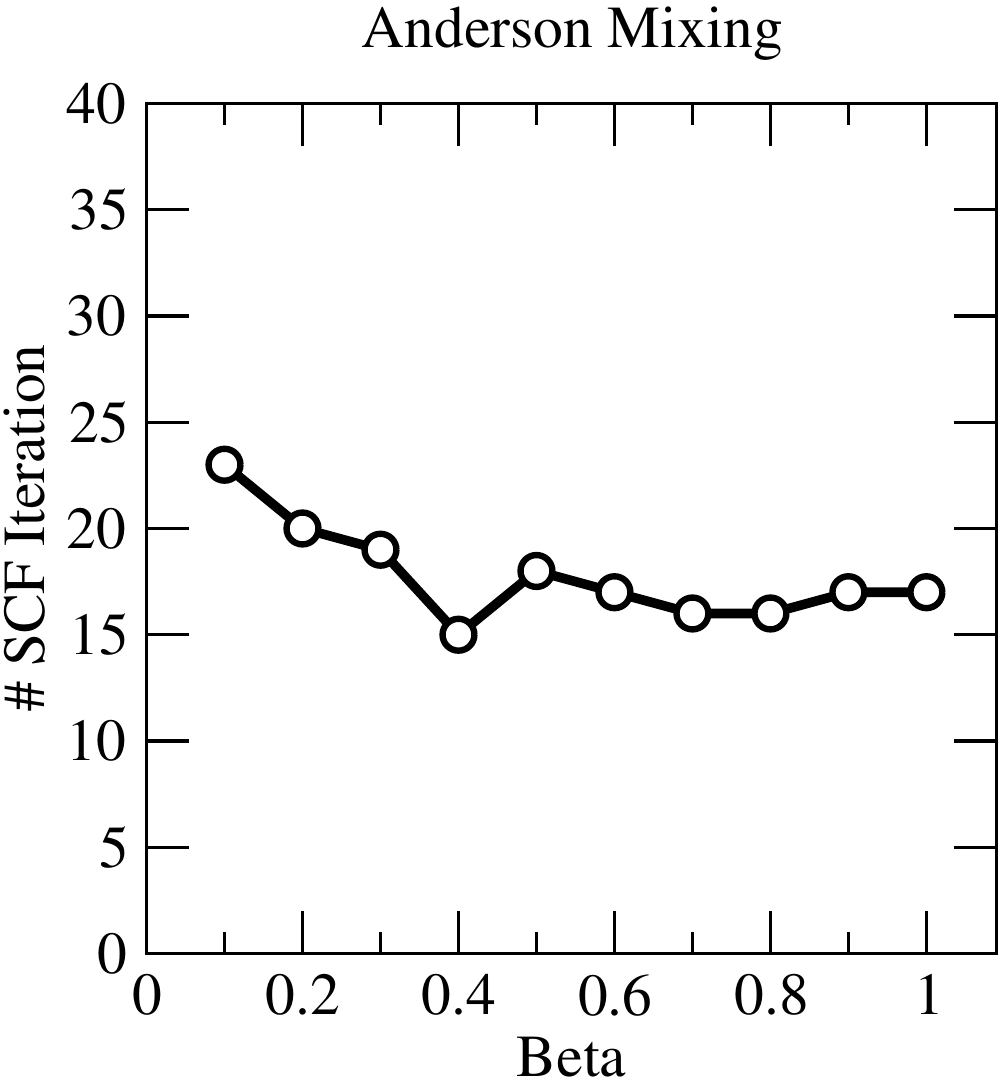}
 \includegraphics[width=0.31\linewidth]{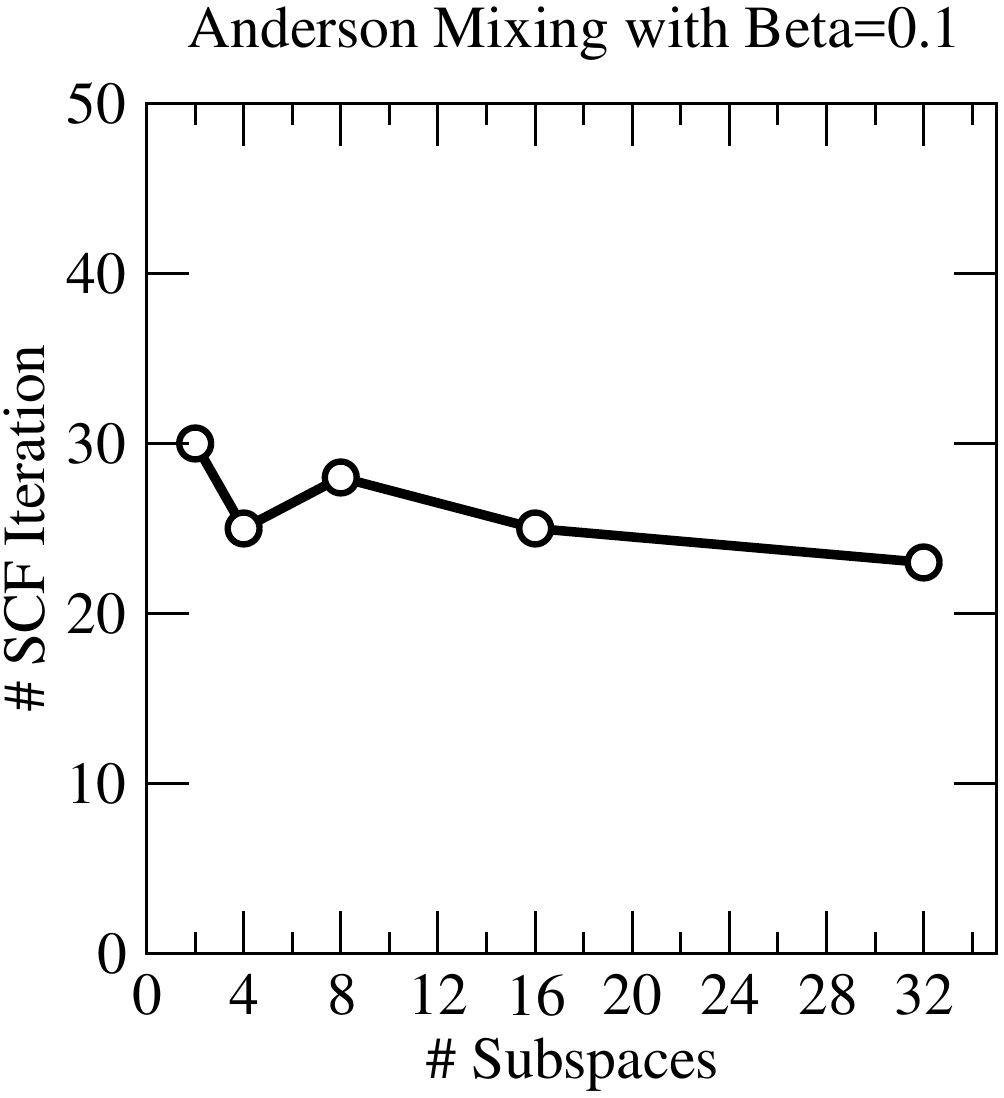}
 \caption{  \label{fig-mixing} The effect of SCF parameters on the convergence of benzene.
The convergence criteria is satisfied when the relative error is below $10^{-10}$ for the total energy. For values of $\beta=0.1$ and $\beta>0.5$ the simple mixing (left) does not converge. With Anderson mixing (middle and right) the total energy converges consistently much faster, and only a few mixing subspaces can be kept in memory.  }
\end{figure}

In NESSIE, the  mixing scheme can take advantage of one important feature of the FEAST eigensolver. 
As the density begins to converge, the previous
eigenvector subspace solution can be used as a very good initial guess for solving the current eigenvalue problem. 
Since the eigenvalue convergence criteria is set to only slightly exceed the current
SCF convergence, FEAST must only perform a single subspace iteration on average
and, with parallelism, solve a single linear system per diagonalization.
%Furthermore, NESSIE can make use of a new algorithm, named NLFEAST (for Non-Linear FEAST)  \cite{gp13},
%that extends the capabilities of the FEAST algorithm to provide a direct solution of the non-linear 'eigenvector' problem (\ref{eq-scf}).
%The approach takes natural advantage of the robustness of the subspace iterations
%procedure of the FEAST algorithm to achieve global convergence (i.e. direct minimization), while
%the non-linearity is only addressed at a level of a reduced system.
%The numerical efficiency of NLFEAST  has been demonstrated in \cite{gp13} using DFT-LDA/all-electron/FEM-P3
%on various molecules, where it was able to 
%outperform the traditional SCF techniques  
%by providing a higher converge rate, convergence
%to the correct solution regardless of the choice of the initial guess, and a significant reduction
%of the eigenvalue solve time in simulations. The approach, however, is currently limited to systems with few hundred
%electrons and further research is needed to improve scalability.

The other major numerical operation in ground-state DFT, is the computation of the Hartree potential - the potential corresponding
to a classical charge distribution - through the solution to the Poisson equation. Once the Dirichlet boundary conditions  have
been determined at the edge the of simulation domain, this amounts to solving a real
symmetric-positive-definite linear system with one right-hand-side.
The Poisson equation gives then rise to a much less expensive linear systems than the ones obtained with the eigenvalue computation.
The computation of the boundary conditions for Poisson uses the integral form of the Poisson equation
and scales as $O(N_s^2/p)$ where $N_s$
is the number of surface nodes (which stays relatively small using a coarse FEM mesh far from the atomistic region), and $p$ is the 
 total number of MPI processes.

Selected DFT/LDA ground-state simulation results for benzene obtained using NESSIE and other all-electron first-principle software,
are reported in Table~\ref{tab-energy}. NESSIE results are in excellent agreement with other approaches.
In addition,  a real-space mesh discretization such as FEM can easily be refined either by adding more local mesh nodes or
by increasing the accuracy of the basis functions. Consequently, the numerical solutions can systematically converge
toward the exact solutions at the level of the physical model (LDA in the example).

\begin{table}[htbp]
  \begin{center}
   \begin{small}
  \begin{tabular}{lccc} 
    Method/Energy(eV) & $\rm E_1$ & $\rm E_{HOMO}$ & $\rm E_{tot}$ \\ \hline\hline
 NWChem 6-311g* \cite{nwchem}&  $-266.35$     &  $-6.40$ & $-6262.27$ \\
 NWChem cc-pvqz \cite{nwchem}&  $-266.41$  &   $-6.52 $  & $-6263.65$ \\ \hline
 P2-FEM \cite{allelec} & $-264.66$   & $-6.54$   &   $-6226.57$       \\
 P3-FEM \cite{allelec} & $-266.38$   & $-6.53$   &   $-6262.57$       \\
 P4-FEM \cite{allelec} & $-266.44$   & $-6.53$   & $-6263.78$        \\ \hline
  FHI-AIMS \cite{allelec,fhi} & $-266.44$   & $-6.53$  &   $-6263.83$ \\ \hline
    NESSIE P2-FEM & $-266.39$ & $-7.04$ & $-6244.45$ \\
    NESSIE P3-FEM & $-266.49$ &  $-6.55$ & $-6263.41$ \\ \hline\hline
  \end{tabular}
  \end{small}
  \end{center}
  \caption{\label{tab-energy} DFT energy results for benzene
    (including the first core eigenvalue, the HOMO level and total energy) obtained using
    different all-electron first-principle software and basis functions, but the same LDA aproach \cite{lda} to model the XC term.
  }
\end{table}

\subsubsection{Discussions on Scalability}

In Table~\ref{tab-energy}, the NESSIE FEM discretization leads to system matrices of size $2,454$ for the atom-centered mesh
(i.e. a single muffin) using P2, or $8,155$ using P3. The interstitial size matrix varies from $20,653$ for P2 to $69,305$ for P3.
These system matrices can  effectively be handled in parallel using the muffin-tin decomposition and spectrum slicing approach
presented in Figures~\ref{fig-muffindd} and \ref{fig-slicing}. 
For example, using only two search intervals (one for the core and one for the valence states, i.e. L1=2 in FEAST), eight contour points per interval (so eight linear systems in total, i.e. L2=8 in FEAST),
a molecule like benzene with twelve atoms (i.e. L3=12 in FEAST) can
effectively scale up to  $2\times 8\times 12=192$ MPI processes on HPC platforms.

In general, the three levels of
parallelism of FEAST can
work together to minimize time spent in all stages of the
algorithm. The third level L3 can be used to reduce the memory
per node and to decrease the solution time of both the linear
system factorization and solve. The second level L2 has close to
ideal scaling, and, if fully utilized, can reduce the algorithmic
complexity to solving a single complex linear system per FEAST iteration. 
The first level L1, in turn, 
allows for the computation of a very large
number of eigenvalues by subdividing the full search interval.
The simulation results obtained in Ref. \cite{pfeast} have demonstrated that the NESSIE's muffin-tin approach
associated with the FEAST eigensolver, is ideally suited for achieving  both strong and weak scalability on high-end HPC platforms.
Some results on weak scalability (i.e. the number of MPI processes increases proportionally with the number of atoms)
are reported in Figure~\ref{fig-sca}. These results outline, in particular, the efficiency of the muffin-tin DD solver presented
in Figure~\ref{fig-muffindd} in comparison with other `black-box' sparse parallel direct solvers.
\begin{figure}[htbp] 
  \begin{small}
\begin{tabular}{l||ccccc}
 \#MPI & 9   &  17 &  25 &  33 &  41 \\
 \hline
  \hline
\#Atoms & 54 & 102 & 150 & 198 & 246 \\ 
System size & 211K & 392K & 575K & 758K & 942K \\ 
\end{tabular}
\end{small}
\centering  
\includegraphics[width=0.7\linewidth]{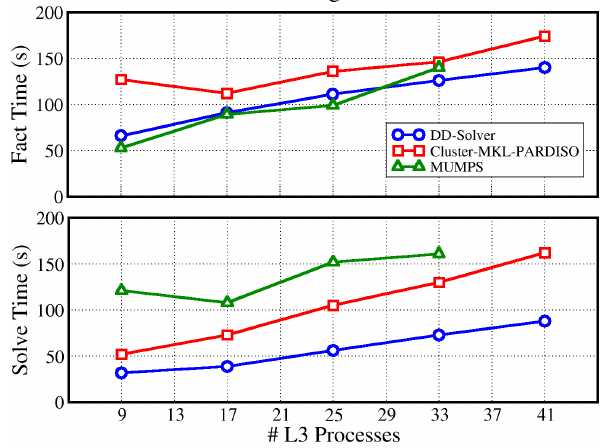}
  \caption{ \label{fig-sca} L3 weak scaling of factorization and 
solve stages for a single FEAST iteration using 
16 contour points (L2=1 here) 
and $600$ right-hand-sides (i.e. search interval with up to $600$ states). The matrix size is increased
 proportionally with the number of MPI processes (using 12 cores per MPI of Haswell E5-2680v3 --
 so 492 cores in total for the 246 atom systems). The results show that the muffin-DD solver in Figure~\ref{fig-muffindd} 
outperforms both MKL-Cluster-Pardiso \cite{intelmkl} and MUMPs solvers \cite{mumps}. % (while also reducing the memory requirement). 
Overall timings can be easily reduced by a factor 16 using L2=16 MPI parallelism  
 (using 7872 cores in total).  Finally, timings can further improve by increasing 
the number of MPI processes for a fixed atom size system (strong scalability).}
\end{figure}

%%%%%%%%%%%%%%%%%%%%%%%%%%%%%%%%%%%%%%%%%%%%%%%%%%%%%%%%%%%%%%%%%%%%
\subsection{Real-time TDDFT Excited States Calculations}
%%%%%%%%%%%%%%%%%%%%%%%%%%%%%%%%%%%%%%%%%%%%%%%%%%%%%%%%%%%%%%%%%%%%

In TDDFT theory, all the occupied $N_e$ ground-state wave functions $\Psi = \{\psi_1 , \psi_2 , \dots, \psi_{N_e} \}$  
solutions of the
Kohn-Sham system for the ground-state problem (\ref{eq-scf}), 
are used as initial conditions for solving a time-dependent 
Schr\"odinger-type equation ($\forall i=1, \dots , N_e)$:
\begin{multline}
\label{eq-tddft}
\imath\hbar\frac{\partial}{\partial t}\psi_i({\bf r},t)=\left[-\frac{\hbar^2}{2m}\nabla^2+v_{KS}[n]({\bf r},t)\right]\psi_i({\bf r},t), \\
 \mbox{with} \quad n({\bf r},t)=2\sum_{i=1}^{N_e}|\psi_i({\bf r},t)|^2,
\end{multline}
where the electron density of the interacting system can then be obtained at any given time from
the time-dependent Kohn-Sham wave functions. In principle, TDDFT can be used to calculate any time dependent observable as
a functional of the electron density. In (\ref{eq-tddft}), the Kohn-Sham potential becomes
a functional of the time-dependent density:
$$v_{KS}(n({\bf r}, t)) = v_{ext} ({\bf r}, t) + v_H(n({\bf r}, t)) + v_{XC} (n({\bf r}, t)),$$
where it is common practice to consider a local dependency on time for the XC potential term  $v_{XC}$ (a.k.a., the adiabatic approximation).

\subsubsection{Real-Time Propagation}
Assuming a constant time step $\Delta_t$, the integral form of (\ref{eq-tddft}) introduces the time-ordered
evolution operator $\hat{U}(t + \Delta_t , t)$, such that:
\begin{multline}
\Psi(t + \Delta t ) = \hat{U}(t + \Delta_t , t)\Psi(t), \\ \mbox{ with } \quad 
\hat{U}(t + \Delta_t , t)= {\cal T}\exp\left\{-\frac{\imath}{\hbar}\int_t^{t+\Delta_t}d\tau H(\tau)\right\}.
\end{multline}

There exists a large number of efficient numerical methods for solving this real-time propagation problem \cite{castro}
which can broadly be classified into two categories:
\begin{itemize}
\item[(i)] PDE-based such as the Crank-Nicolson (CN) scheme \cite{CN} where
  \begin{multline}
     \label{eq-cn}
  \hat{U}(t+\Delta_t,t)=\left[1+\frac{i}{2}\Delta_t{H}(t+\Delta_t/2)\right]^{-1}\times\\
  \left[1-\frac{i}{2}\Delta_t{H}(t+\Delta_t/2)\right].
  \end{multline}
This is an implicit scheme that requires solving a linear system at each-time step.  
\item[(ii)] Integral-based that acts directly on the evolution operator, such the mid-point exponential rule i.e.
  \begin{equation}
    \label{eq-exp}
  \hat{U} (t + \Delta_t , t) = \exp\{-\imath\Delta_t H(t+\Delta_t/2)\}.
\end{equation}
This represents the starting point for splitting methods \cite{split}, Magnus expansion \cite{magnus} or other spectral decompositions \cite{cp2010,varga}.  
\end{itemize}
If $\Delta_t$ is small enough, one can usually assume that $H(t+\Delta_t/2)\simeq H(t)$. Alternatively,
corrector-predictor schemes can be used to evaluate the Hamiltonian  at $(t + \Delta_t/2)$. 
The non-linear nature of the time propagation arises from the Kohn-Sham $V_{KS}$ potential that needs
to be reevaluated at each time step to form the new Hamiltonian. 
The Hartree potential $V_H$ and exchange
and correlation potential $V_{XC}$ are both functionals of the electron density. 
The time dependent Hartree potential is solution of the Poisson equation, while the XC term must be computed with
time instantaneous electron density (using the adiabatic approximation) and the same approximation used in ground-state calculations (such as LDA, GGA, etc.).

The CN scheme for TDDFT is particularly effective within the NESSIE framework, since the linear system solves along
the time steps in (\ref{eq-cn}) can take advantage of the highly scalable muffin-tin real-space domain decomposition solver shown in Figure~\ref{fig-muffindd}.
As a result, the approach can be parallelized at two different MPI levels: (i)~by propagating independent
chunks of the occupied wavefunctions along $\Delta_t$; and (ii)~by solving the linear system in parallel.

Spectral-based schemes for the integral-approach (\ref{eq-exp}) are known to be robust and accurate 
 permitting larger
 time steps than other usual integration schemes. %  (such as splitting techniques).
% Similarly to the CN scheme, the latter require small time intervals to guarantee accuracy and
% accommodate the high spatial resolution (typically, $\Delta_t \sim 5as$,  for performing UV-ViS spectroscopy).
% Although spectral approach permits to consider much larger time intervals (typically $\Delta_t \sim 30as$),
% they are rarely used in practice for large-scale simulations
They are, however, rarely used in practice for large-scale simulations
 since 
a direct diagonalization of the evolution operator (\ref{eq-exp}) would require solving hundreds
to thousands eigenvalue problems along the time-domain (i.e. one large-scale eigenvalue problem per
time step). In addition to CN, NESSIE includes an efficient spectral-based approach which relies on the efficiency
of the
FEAST eigensolver \cite{cp2010}.
First, good approximations of the exponential in (\ref{eq-exp}) 
can be obtained with a partial spectral decomposition using an eigenvector subspace four to five times
the number of  propagated states ${\psi_j}$. Since the latter are
low-energy states,  this truncated spectral basis is typically sufficient to accurately expand the solutions.
%First, since the $N_e$ propagated states ${\psi_j}$ are 
%low-energy states, good approximations of the exponential in (\ref{eq-exp}) 
%can be obtained by using a partial spectral decomposition, where one considers a basis of eigenpairs of size $M$
%much smaller than the  Hamiltonian system size $N$, but greater than $N_e$ (i.e. $N_e < M << N$),  and the choice
%of $M \approx 4N_e$  is typically sufficient to accurately expand the solutions.
Second, FEAST can reuse the eigenvector subspace computed in the current time-step as very good initial guess for the next one.
As a result, only
 one or two subspace iterations are usually sufficient to obtain  convergence.
While the linear systems arising in CN need to be solved one
after another along small time-intervals, a parallel FEAST implementation 
permits the solution of a single linear system by larger time intervals.
Although the FEAST linear system is notably more computationally demanding i.e. including a lot of extended states,
linear parallel scalability
can still be naturally achieved using multiple search intervals and more parallel computing power. 
It is worth mentioning that the same strategy would not be possible with the techniques used in 
TDDFT linear response theory in frequency domain (e.g. using Casida equation \cite{TDDFTbook}),
 where the demand in extended states is even higher and represents the bottleneck of their cubic arithmetic complexity.
Consequently,  if one can keep up with the demand in
 parallel computing power, % (ultimately 
%using petascale or exascale architectures),
 direct diagonalizations for the real-time TDDFT formalism 
using FEAST become a viable high-performance alternative to other schemes, potentially capable of both 
higher-scalability and better accuracy for obtaining linear and non-linear responses.

\subsubsection{Dipole-time Response}
%Similarly to DFT, all observables in TDDFT must be expressed in terms of the electron
%density, which is now time dependent.
In principle,
all time dependent observables are functionals of the density, however, in practice the
functional form is rarely known. A very useful case of TDDFT is related to spectroscopy, where
the absorption and emission spectrum corresponding to electronic excitations can be
 derived directly from the induced dipole moment $d(t)$.
 The latter is related to the response of the system to an applied electric field $E$:
\begin{equation}
\label{eq-dipole1}
d(t) = \int _{-\infty} ^t \alpha(t-t^\prime)E(t^\prime) dt^\prime,
\end{equation}
where  $\alpha$ is defined as the dynamic polarizability. In general, $E$ and
$d$ are vectors quantities with $x$, $y$ and $z$ components and $\alpha$ is a tensor.
In computational spectroscopy, one often considers the response of the system in a given direction $\mu$ (e.g. $x$, $y$ or $z$)
associated with an excitation polarized  in the same direction.
The induced dipole moment in (\ref{eq-dipole1}) can be calculated as 
a measure of how far the electron density $n(r,t)$ has moved away from its ground-state
value $n_0$  along a given direction $\mu$ \cite{octopus,andrade,takimoto}:
\begin{equation}
  \label{eq-dt}
d _{\mu} (t) = -q\int _\Omega (\mu - r_0) \left( n(r,t)-n_0(r) \right) dr,
\end{equation}
where $r_0$ stands for the molecular center of mass. As a result, for an isotropic material, it is possible to compute
the dynamic polarizability by inverting (\ref{eq-dipole1}). In frequency domain (after Fourier transform), the expression becomes: 
\begin{equation}
\label{eq-alpha}
\alpha(\omega) = \frac{d(\omega)}{E(\omega)}.
\end{equation}
In practice, it is necessary to introduce artificial damping signal into the computed dipole moment before taking its Fourier transform:
\begin{equation}
  \label{eq-domega}
d(\omega) = \int _0 ^T \left( d(t) \times e^{-\gamma t} \right) e^{i\omega t} dt,
\end{equation}
where $\gamma$ is a damping coefficient. This damping term is used to mimic the system relaxation effect
since the TDDFT simulations presented here do not explicitly account for energy dissipation
(i.e. a system would physically emit energy as photons and relax back to the ground-state after being excited).

In time-dependent simulations, any external electric field $E(t)$ may be considered. It is common practice, however, to
 either use a step potential  $u(t)$ or an impulse excitation $\delta(t)$ both along a given direction \cite{YB2}.
 Table~\ref{tab-Ew} presents the resulting expressions for the dynamic polarizability $\alpha(\omega)$ after
Fourier transforms of these particular electric fields.
\begin{table}[htbp]
$$
\begin{array}{cc} 
	E(t)                  & \alpha(\omega) \\
\hline
\hline \\
       E_0 \times \delta(t)  &   -d(\omega)/E_0  \\[5pt]
       E_0 \times u(t)      &   -{i \omega} \times  d(\omega)/E_0 \\ \hline\hline 
\end{array}
$$
\caption{ 
\label{tab-Ew}
Polarizability for excitations in the form of an impulse potential and a step potential
($E_0$ stands for the amplitude of the electric field).}
\end{table}

Finally, the imaginary part of the dynamic polarizability provides the photo-absorption cross section \cite{astapenko2013interaction}, 
which is related the probability that a photon passing through the atomistic system is absorbed.
A measure of the strength of this interaction (a.k.a., the oscillator strength) can be computed from the polarizability as follows \cite{hilborn1982einstein,chen13}:
\begin{equation}
\label{eq-sigma}
\sigma(\omega) = \frac{4 \pi \omega}{c} \Im(\alpha(\omega)).
\end{equation}
Once the oscillator strength $\omega$ is plotted in function of the frequency
 $\omega$ (or equivalently the absorption energy), it provides the absorption spectrum of the system. 
As an example, Figure~\ref{fig-dipole}
shows the variations of the induced dipole moment for benzene obtained after three distinct
impulse excitations polarized in the $x$, $y$ and $z$ directions, as well as the corresponding three absorption spectra.

\begin{figure*}[htb]
  \centering
 \includegraphics[height=0.3\linewidth]{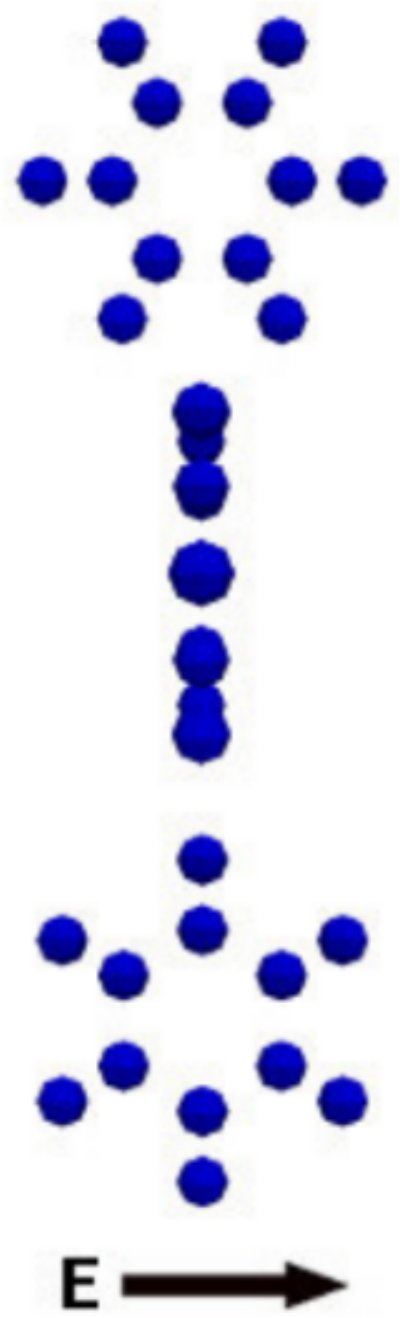}
 \includegraphics[height=0.3\linewidth]{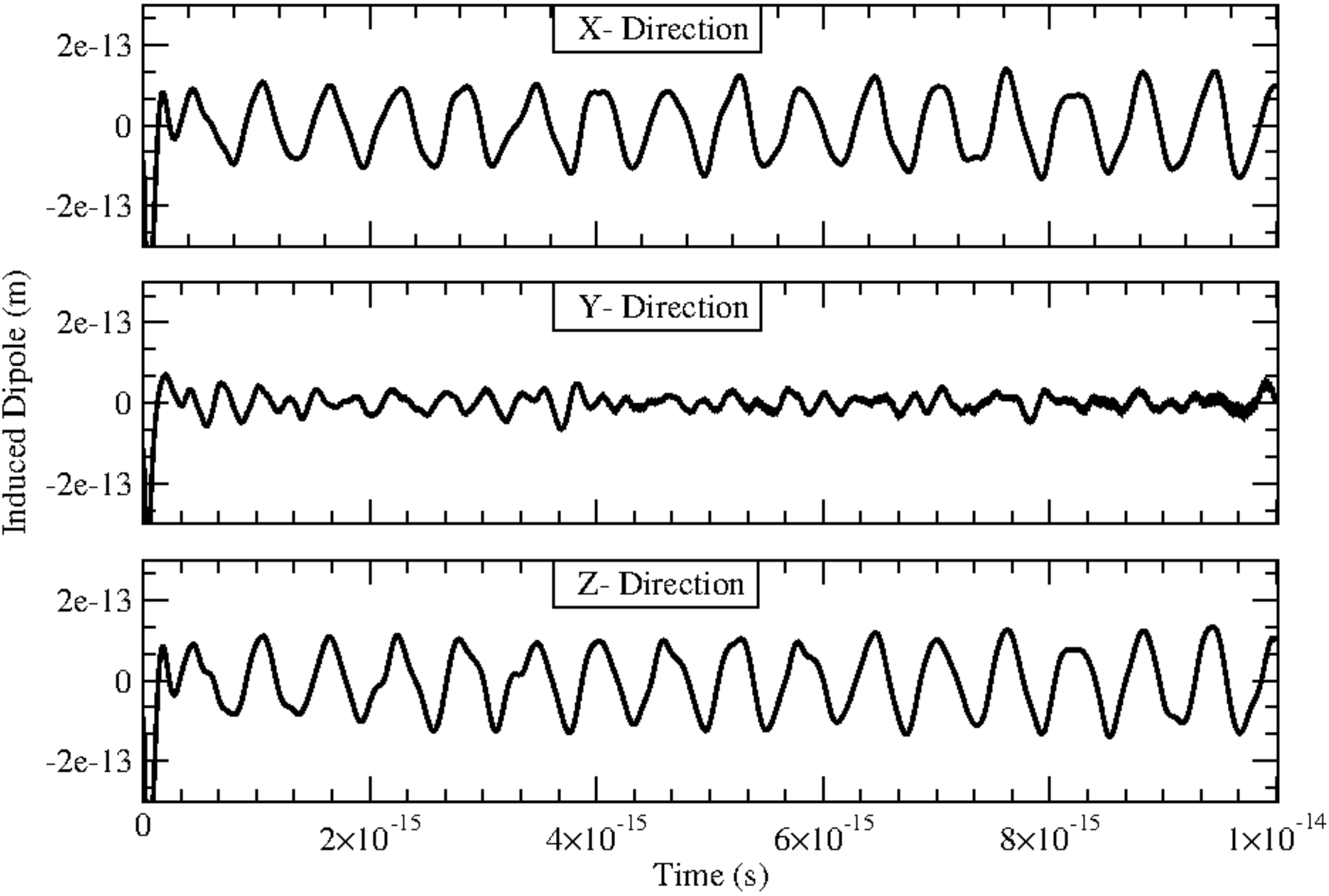}
 \includegraphics[height=0.3\linewidth]{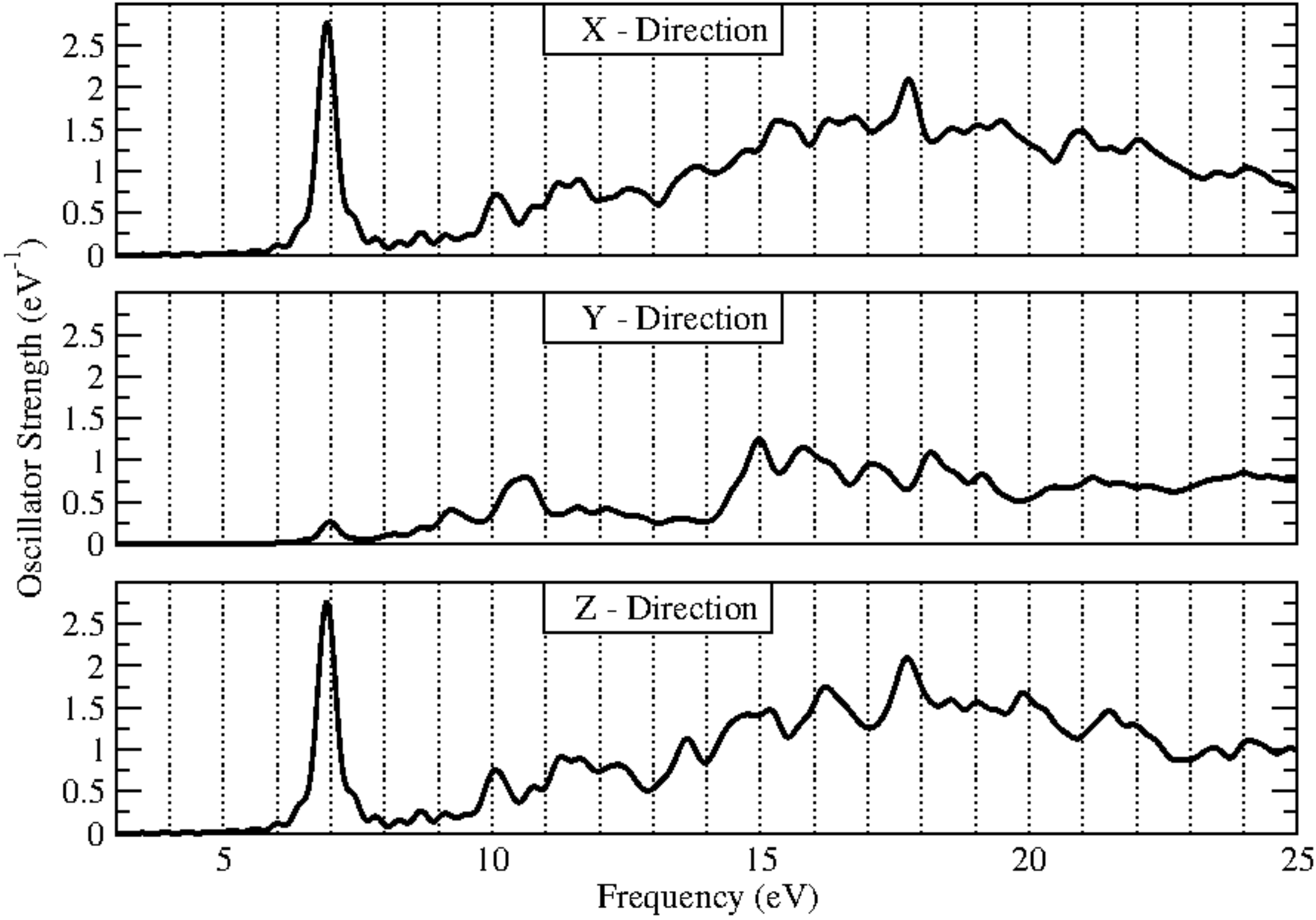}
 \caption{  \label{fig-dipole}
   The left plots represent the relative orientation of the benzene molecule if the applied electric 
   field was polarized from left to right. The middle plots show the induced dipole moment for benzene after an impulse excitation polarized in the $x$ (top), $y$ (middle) and $z$ (bottom) direction (time steps of 5as are used).
   The corresponding absorption spectra are shown in the right plots. 
 }
\end{figure*}

\subsubsection{Resonances and response density}

The peaks in the absorption spectrum correspond to specific quantum many-body excitations (such as plasmon, band-band, etc.).
The electron dynamics for a specific peak can be investigated further by computing and then visualizing the response density in 4D.
Such simulations aim at providing more details on the electron
dynamics of the particular resonances with relevant information about their nature. 
The response density $\delta n(\omega,r)$ is the change in electron density due to an excitation at frequency $\omega$
(i.e. charge oscillations at $\omega$). 
One possible way to visualize $\delta n(\omega,r)$ is  by applying a sinusoidal excitation at a given frequency of interest,
waiting for the induced dipole moment to reach a steady state where it oscillates at $\omega$, and plotting the 4D data
when the dipole reaches  a maximum and a minimum \cite{ps14}.
Another more efficient approach consists of computing $\delta n(\omega,r)$ directly following
 the same procedure used for deriving the dipole moment in (\ref{eq-dt})
and (\ref{eq-domega}), which leads to:
\begin{equation}
  \label{eq-nw}
\delta n(r,\omega) = \int _0 ^T \left(n(r,t) - n_0(r) \right) e^{- \gamma t}) e^{i \omega t } dt.
\end{equation}
In practice, there is no need to store all the $n(r,t)$ functions that would be too prohibitive. 
Once the peaks/resonances of interests (i.e. $\{\omega_j\}$) have been identified in
the absorption spectrum (for a given polarized excitation),
one can proceed by running a new time-dependent TDDFT calculation to compute
the response density $\delta n(r,\omega_j)$ (all frequency $\{\omega_j\}$ at once) using
an on-the-fly Fourier transform of the time varying electron density.
Results from this approach are shown in Figure~\ref{fig-peaks} for few selected peaks in the absorption spectrum of benzene
when the molecule is excited with an impulse electric field polarized in the $z$ direction (as shown in Figure~\ref{fig-dipole}).

\begin{figure}[htbp]
  \begin{center}
\begin{minipage}{.24\linewidth}
  \centering
  \includegraphics[height=\linewidth,angle=-90]{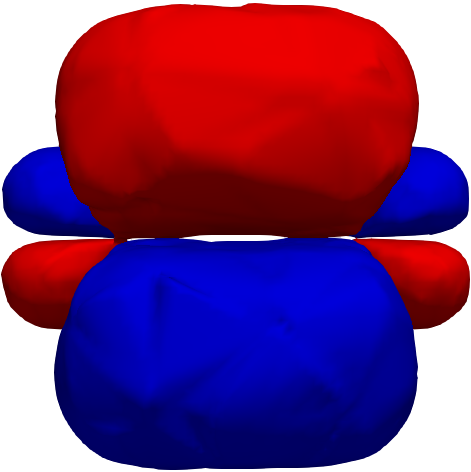}
\end{minipage}
\begin{minipage}{.24\linewidth}
  \centering
  \includegraphics[height=\linewidth,angle=-90]{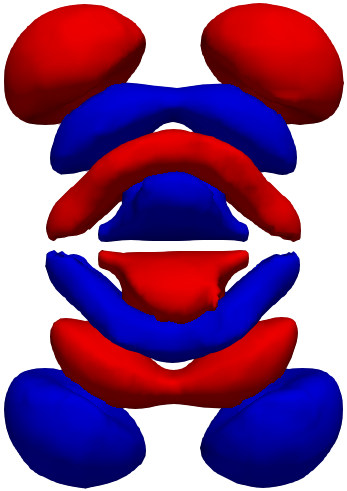}
\end{minipage}
\begin{minipage}{.24\linewidth}
  \centering
  \includegraphics[height=\linewidth,angle=-90]{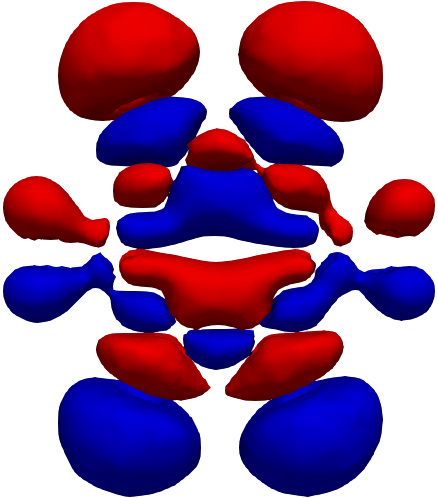}
\end{minipage}
\begin{minipage}{.24\linewidth}
  \centering
  \includegraphics[height=\linewidth,angle=-90]{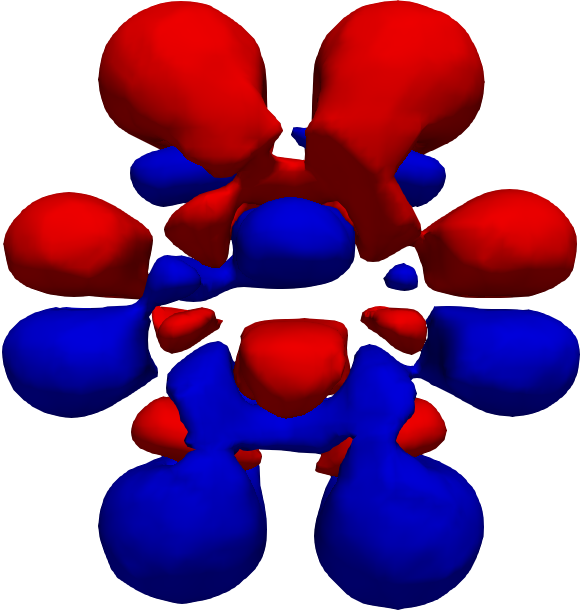}
\end{minipage}
\end{center}
  \caption{\label{fig-peaks}
    4D isosurface plots of the response electron density for four different peaks chosen from the absorption spectrum of $z$-polarized excited
    benzene (electric field going from left to right).
    From left to right, four frequency $\omega$ values are represented:
     $6.91$eV, $10.04$eV, $13.61$eV, and $16.20$eV. 
          }
\end{figure}

\subsubsection{Discussion on Accuracy and Reliability}
The direction independent absorption spectrum which is computed as the average of the spectra in $x$, $y$ and $z$ directions, can
be directly and {\em quantitatively} compared with the experimental data, if available.
Figure~\ref{fig-exp} compares the experimental absorption spectra of various molecules with the NESSIE's first-principle simulation results  (obtained at T=0K).
In general, the TDDFT simulation results compare remarkably well with experimental data for a large number of atomistic systems.
\begin{figure*}[htb]
  \centering
 \includegraphics[height=0.25\linewidth]{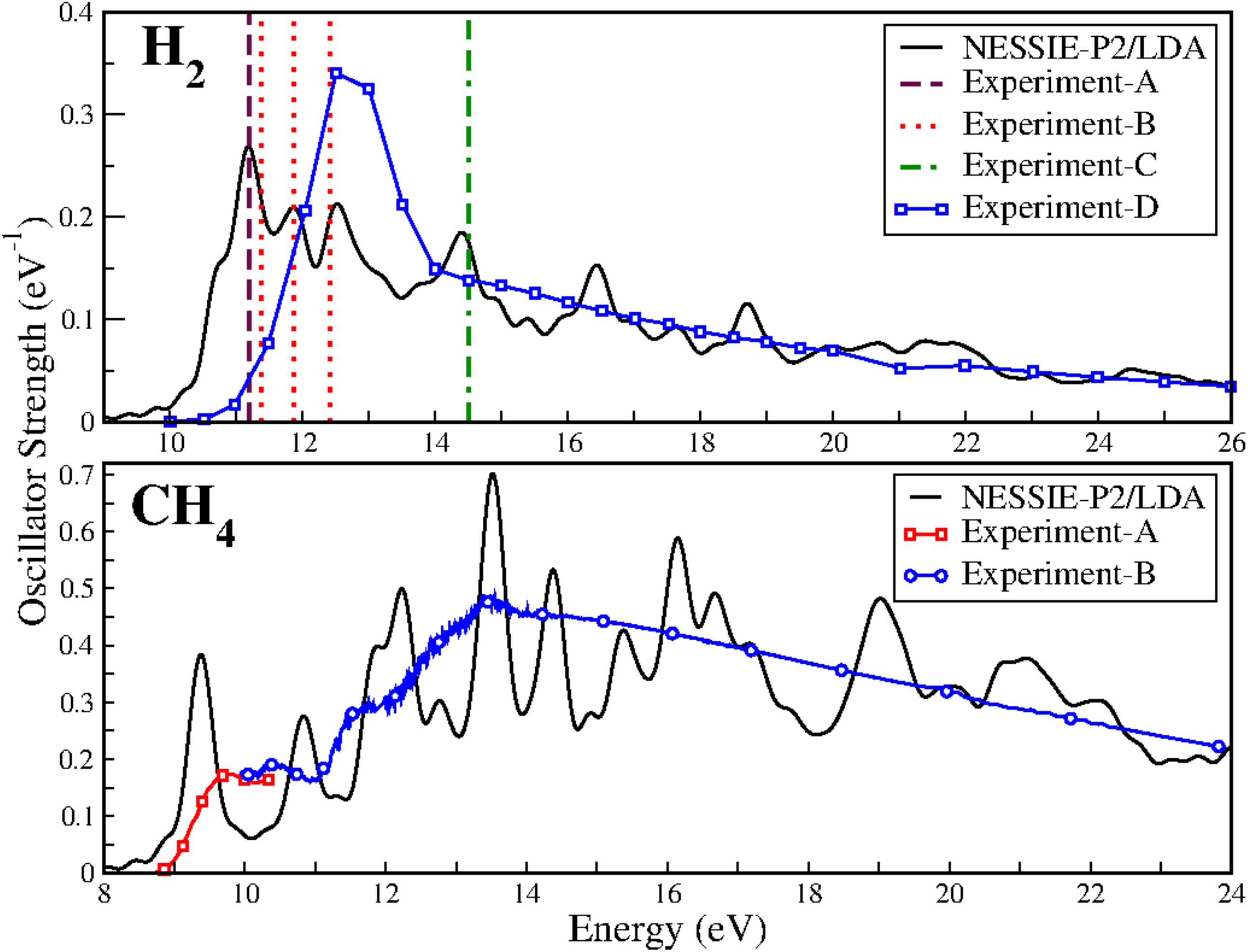}
 \includegraphics[height=0.25\linewidth]{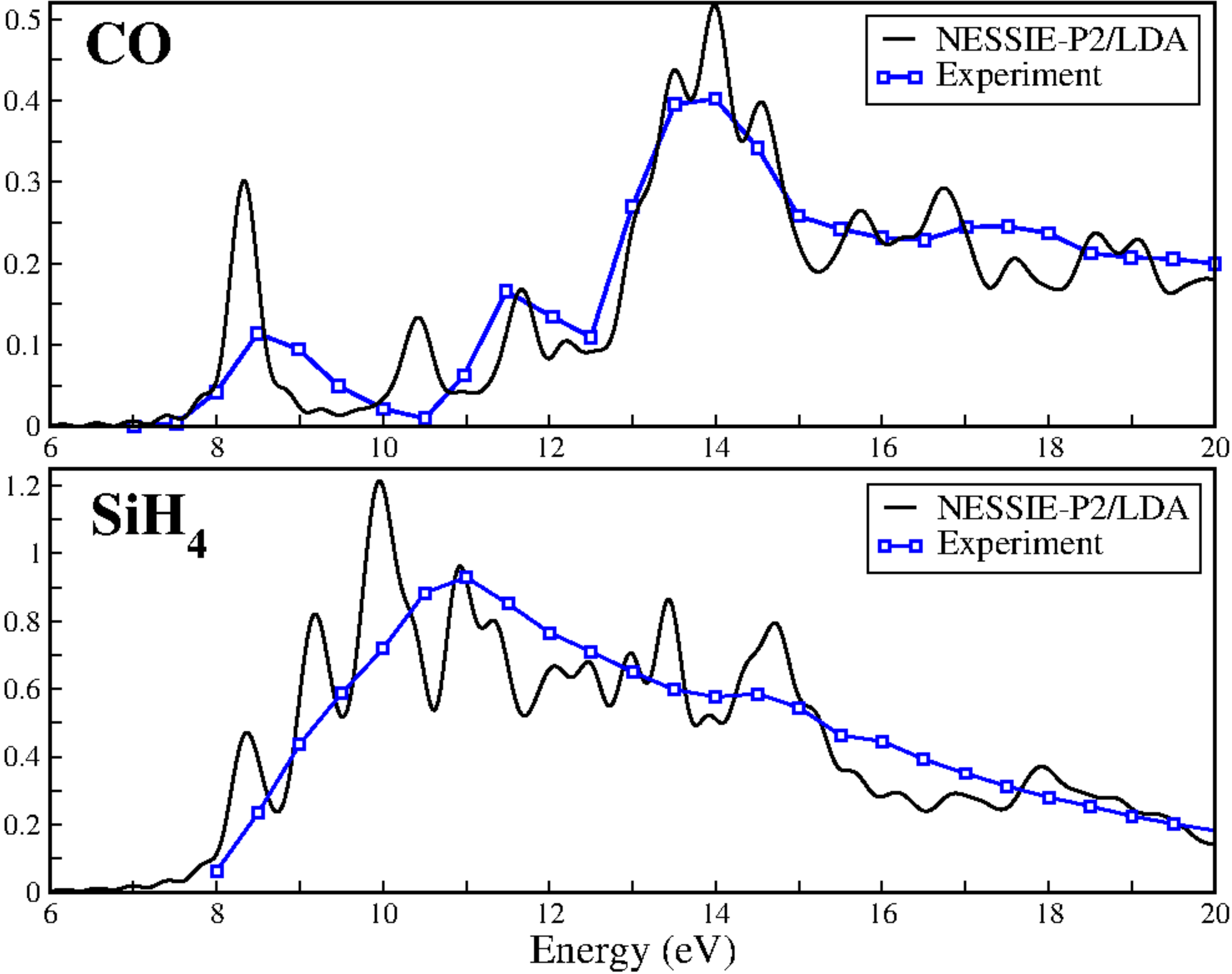}
 \includegraphics[height=0.25\linewidth]{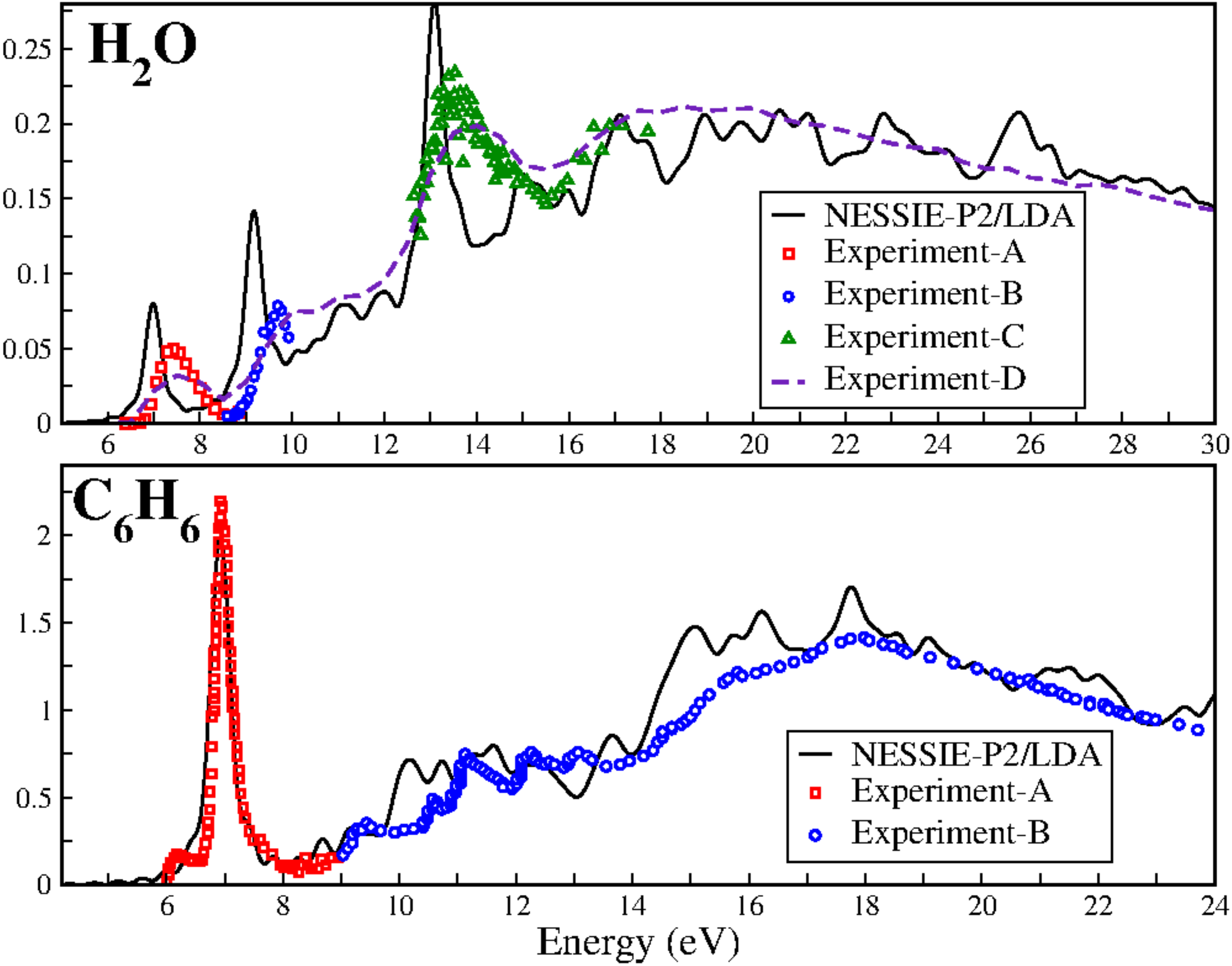}
 \caption{  \label{fig-exp}
   Comparison between the computed absorption spectrum for various molecules using NESSIE (with ALDA and P2 FEM) and multiple set
   of experimental values for $\rm H_2$ (Exp-A \cite{h2a}, B \cite{h2b} C \cite{h2c} and D \cite{h2d}), $\rm CO$ \cite{co}, $\rm H_2O$ 
   (Exp-A \cite{h2oa}, B \cite{h2ob} C \cite{h2oc} and
   D \cite{h2od}),  $\rm CH_4$ 
   (Exp-A \cite{ch4a}, and B \cite{ch4b}), $\rm SiH_4$ \cite{sih4},
   and $\rm C_6H_6$  (Exp-A \cite{c6h6a}, and B \cite{c6h6b})}
\end{figure*}

While the choice of the XC functional can significantly impact the reliability of the DFT ground-state results,
a simple adiabatic LDA (ALDA) approximation for TDDFT appears to be sufficient for a wide variety of systems.
Using NESSIE, it is also interesting to note that the
choice of P2 vs P3 FEM basis functions, does have only
a minimal impact on the accuracy of the absorption spectrum \cite{james}.
This is clearly not the case for DFT ground-state calculations
as reported in Table~\ref{tab-energy}, where good accuracy would require
an appropriate level of refinement for FEM (using at least
cubic P3 FEM). As a result, the real-time TDDFT framework appears
to be resilient to some approximations (such as the choice of XC term, or basis functions) as long as they are not too far off
and stay consistent throughout the time propagation.
%It can also explain why the choice of LDA
%is not a limiting factor of the physical model for many
% molecules as long as we are looking at linear responses (e.g. using a  weak initial impulse). 
%The small quantitative differences with experimental data should rather be attributed to  the adiabatic approximation
% and the fact that the model
%does not account for the memory effects.

Beside offering the opportunity
to perform X-Ray spectroscopy, there are many other advantages for considering a full real-space all-electron treatment in simulations.
In contrast to other approaches, a full-core real-space potential offers numerical consistency while performing simulations in time-domain.
Transferability issues are indeed likely happen with the use of
pseudopotentials that are generated for time-independent calculations, or in turn, with the use of LCAO basis
which cannot offer  the same reliability to capture both confined and extended states (although LCAO bases 
can be ``augmented'' in time-dependent simulations).
%In comparison with other real-space pseudopotential code
%such as the Octopus package \cite{Octopus0}  applied to various organic molecules
%\cite{co1,takimoto,ch4}, all-electron simulations present
%more favorable agreements with the experimental data.
Comparisons between NESSIE and LCAO approaches are reported in Table~\ref{tab-peaks}.
Although NESSIE is using both a low level of real-space approximation (P2-FEM) and
a rather simple XC term (LDA), the results compare relatively well with the experimental data.
These results are actually much better than the ones obtained with the NWChem software using
the ccpvtz basis and more advanced XC term (B3LYP)  (except for the $\rm CO$ molecule).

\begin{table}[htbp]
  \begin{center}
   \begin{small}
     \begin{tabular}{lcccc}
              & $\rm H_2$ &  $\rm CH_4$ &  $\rm CO$ & $\rm C_6H_6$ \\ \hline\hline
  NWChem-ccpvtz/LDA  & $12.31$ & $10.29$ & $8.28$ &  $7.10$ \\
  NWChem-ccpvtz/B3LYP & $12.90$ & $10.75$ & $8.55$ & $7.18$  \\
  NESSIE-P2/LDA & $11.19$ & $9.40$ & $8.40$ & $6.90$ \\
  Experiment & $11.19$ & $9.70$ & $8.55$ & $6.90$ \\ \hline\hline
  \end{tabular}
  \end{small}
  \end{center}
  \caption{\label{tab-peaks} 
    Comparison of Real-Time TDDFT NESSIE and NWChem simulation results, with
    the experimental data for the lowest excitation energies (in eV), as reported in \cite{lopata}.}
\end{table}

\subsubsection{Discussion on large-scale real-time TDDFT simulations} 
The computing challenges end up being very similar between
ground-state DFT and excited-state real-time TDDFT calculations. There are only two main operations to consider:
(i)~solving a Hamiltonian linear system with multiple right hand sides using in particular the muffin-tin technique in Figure~\ref{fig-muffindd};
and (ii)~solving a linear system for the Poisson equation (using local XC). For DFT, these two steps have
to be repeated self-consistently until convergence, while for TDDFT, they need to be repeated at
each time step of the time-propagation (using either a Crank-Nicolson or a spectral decomposition
scheme). In comparison to other approaches, the muffin-tin solver has demonstrated great efficiency to achieve
strong and weak scalability \cite{pfeast} (see Figure~\ref{fig-sca}).
The scalability bottleneck of the muffin-tin solver would eventually come from solving the interstitial
Hamiltonian system in parallel using MPI (Schur complement in step 2 of Figure~\ref{fig-muffindd}).
In practice, our strong and weak scalability results show that up to one thousand atoms
(corresponding to $\sim 1$M size interstitial matrix), this system can be efficiently solved using a standard direct
'black-box' parallel sparse system solver (such as cluster MKL-PARDISO \cite{intelmkl}). Reaching
the milestone of ten thousand atoms and beyond, is still the subject of active research efforts that
investigate  new directions in numerical linear algebra such as the use of
 hybrid parallel solvers with customized low-communication preconditioners.

\section{NanoPlasmonic Applications}\label{sec-4}

Nanoplasmonics is a field that has grown rapidly in the last few 
 years  \cite{y1,y2}, 
and it already   offers   numerous   applications   to
electronics and photonics. In the visible and near IR  range,  nano-antennas
\cite{y3}  and  nanoparticles  have  provided  drastically  enhanced  coupling  to
electromagnetic  waves  \cite{y4,y5}.
A 2008 comment  in  Nature
Nanotechnology \cite{y3} stated ``{\em molecular  components  promise  to  revolutionize
the electronics industry, but the  vision  of  devices  built  from  quantum
wires and  other  nanostructures  remains  beyond  present  day  technology.
Making such devices will require an  extremely  detailed  knowledge  of  the
properties of these components such as  the  dynamics  of  charge  carriers,
electron spins and various  excitations,  on  nanometer  length  scales  and
subpicosecond timescales at very high (up to THz)  frequencies}''. 
  Reference \cite{y3} points out that single-wall carbon nanotube (SWCNT) resonators would constitute a  unique  THz  
ultra-compact circuit element, which might for example be used to  control  a  THz
oscillator source.  Similar  opportunities  exist  in  the  IR  and  visible
ranges.  One  concludes  that  discovery  of  new  plasmonic
materials is mandatory for  the  future  expansion  of  the  field  of  nanoplasmonics.

In Ref. \cite{ps14}, NESSIE has been used to 
 provide evidence of the plasmon resonances (collective electron excitations) in a number
 of representative short 1D finite carbon-based nanostructures using  real-time TDDFT simulations.
The simulated systems ranged from small molecules
such as $\rm C_2H_2$ to various carbon nanostructures that are equivalent to 1D conductors
with finite lengths, including: carbon chains, narrow armchair 
and zigzag graphene nanoribbons (i.e. acenes and PPP), and short carbon nanotubes (CNT).
NESSIE all-electron TDDFT/ALDA's model was able to accurately capture the bright components of the spectra 
which account for the plasmonic excitation.
The chief  signature  of  1D  plasmons  is  a  high-frequency excitation that is inversely proportional to  the  length  of  the
conductor. In particular, it was shown
that metallic 1D CNTs can  be  well
described with the Tomonaga-Luttinger
theory \cite{ps14,y6,y7,y8,y9,y12}. 
The plasmon velocity is expected to reach an asymptotic value
(up to 3 to 5 times the single particle Fermi velocity)
when the simulations are extended to tens of unit cells, such as
very long CNT that become relevant for THz spectroscopy.

Since the reported preliminary work on short CNT \cite{ps14} (about 5 unit-cells), NESSIE has been upgraded to simulate large-scale
atomistic systems by taking advantage of new high performance computing techniques such as the muffin-DD solver presented and discussed
in Figures~\ref{fig-mesh} and \ref{fig-muffindd}. This all-electron real-space and real-time TDDFT framework is now capable to
simulate very large structures (up to tens of unit cells for CNT -- from hundred atoms to few thousands),
and lead to more relevant predicted data of the plasmonic effects for 1D systems.
As an example, Table~\ref{tab-cnts} summarizes the main NESSIE parameters and simulation results
while considering increasingly longer (3,3)-CNTs.
These results show that the position of the plasmon excitation peak (lowest excitation energy/frequency i.e. $E_p$, $f_p$)
keeps shifting with longer CNTs. The corresponding absorption spectra are provided in Figure~\ref{fig-cnts}.

\begin{table}[htb]
 \begin{center}
 \begin{tabular}{clcccc}
   & & 5-CNT & 10-CNT & 20-CNT & 40-CNT \\ \hline\hline
   \multirow{3}{*}{\rotatebox[origin=c]{90}{\parbox[c]{1cm}{\centering System}}}
   & Length (nm) & $1.26$ & $2.49$ & $4.99$ & $9.99$\\
 & \#Atoms & $78$&$138$&$258$&$498$\\
   & \#Electrons &$408$&$768$&$1488$&$2928$\\
   \hline
\multirow{3}{*}{\rotatebox[origin=c]{90}{\parbox[c]{1cm}{\centering Mesh}}}
 & size muffin & $2065$ & $2065$ & $2065$ & $2065$\\
& size inters. & $149499$&$126431$&$233696$&$446559$\\
& size total &$302925$&$397877$&$741182$&$1426125$\\ \hline
\multirow{3}{*}{\rotatebox[origin=c]{90}{\parbox[c]{1cm}{\centering DFT}}}
 & $\rm E_1$ (eV)& $-267.73$ & $-268.05$ & $-268.07$ & $-268.28$\\
& $\rm E_{HOMO}$ (eV)& $-4.996$&$-4.823$&$-5.000$&$-4.955$\\
& $\rm E_{tot}$ (eV)&$-67713$&$-128825$&$-251351$&$-496359$\\ \hline
\multirow{3}{*}{\rotatebox[origin=c]{90}{\parbox[c]{1cm}{\centering RT-CN}}}
 & $\Delta_t$ (fs)& $0.01$ & $0.01$ & $0.01$ & $0.01$\\
& Total T (fs)& $15$&$30$&$45$&$65$\\
& \#Time-steps&$1500$&$3000$&$4500$&$6500$\\ \hline
\multirow{3}{*}{\rotatebox[origin=c]{90}{\parbox[c]{1cm}{\centering TDDFT}}}
 & ${\rm E_{p}}$ (eV)& $1.62$ & $1.24$ & $0.82$ & $0.525$\\
& $\rm f_{p}$ (THz)& $394.13$&$299.83$&$198.27$&$126.94$\\
& $\rm v_{p}$ ($10^6$m/s)&$0.98$&$1.49$&$1.98$&$2.54$\\ 
 \hline \hline
 \end{tabular}
 \end{center}
  \caption{\label{tab-cnts}
   NESSIE's main parameters and simulation results obtained
using the all-electron/DFT/TDDFT/ALDA framework applied to increasingly longer (3,3)-CNTs.
The real-space mesh uses a P2-FEM discretization while the real-time approach uses a CN (RT-CN) propagation scheme. The simulations
were executed on Xsede-Comet \cite{xsede}. }
\end{table}

\begin{figure}[htb]
  \centering
 \includegraphics[width=\linewidth]{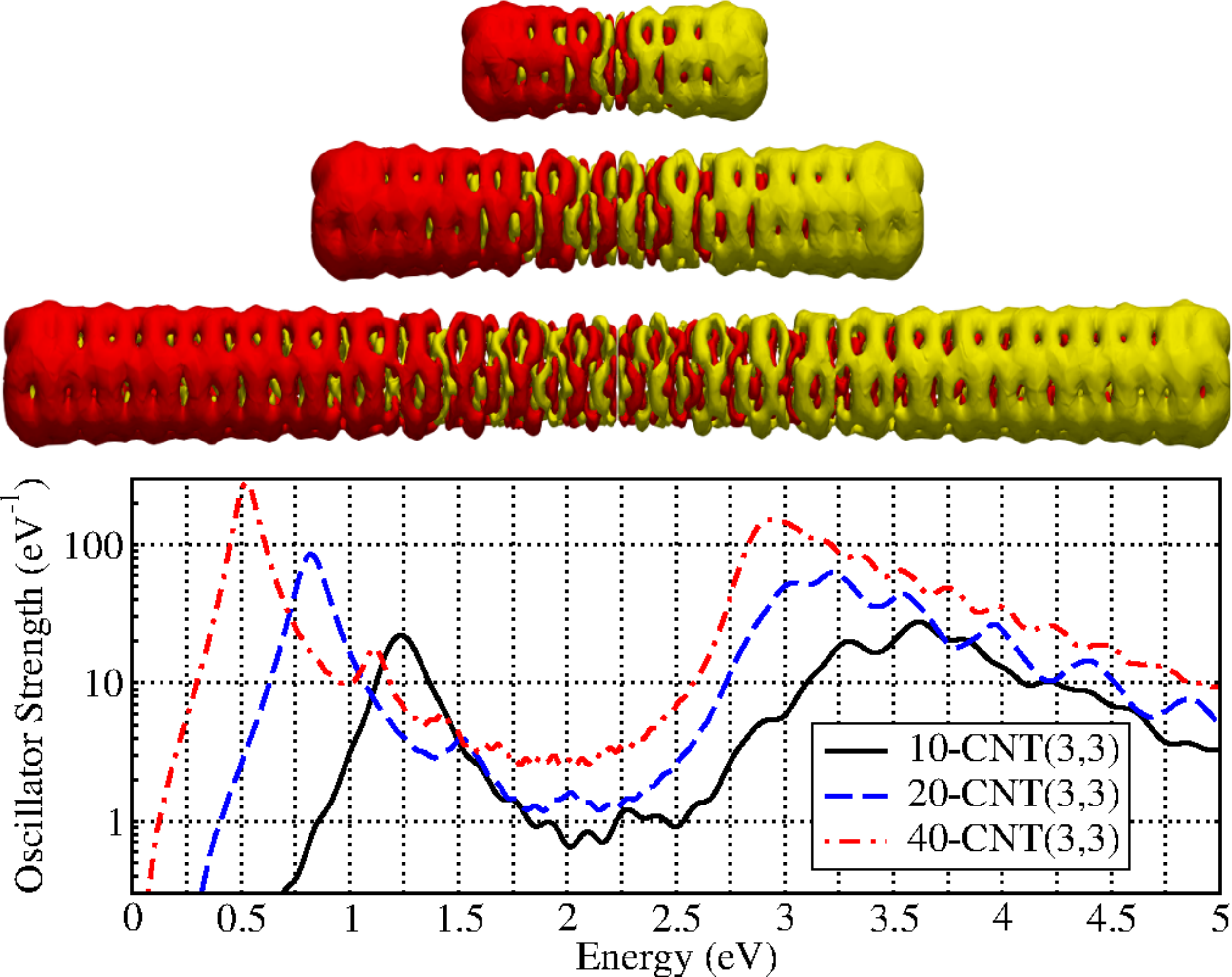}
 \caption{  \label{fig-cnts}
   Computed absorption spectra for three (3,3)-CNTS presented in Table~\ref{tab-cnts} (10, 20 and 40 units cells).
   The selected energy range outlines the lowest excitation peak which keeps shifting left (red-shift)
   with longer nanostructures. The corresponding 4D isosurface plots of the response electron density (on top)
   provide visual information about the dynamic of these plasmonic excitations.
 }
\end{figure}

In Table~\ref{tab-cnts}, the plasmon velocity is obtained with the reasonable
assumption that the plasmon (collective electron cloud) must travel back and forth the full length $L$ of the nanotube
to complete a single oscillation (i.e. $v_p=2Lf_p$). This is further supported by the 4-dimensional isosurface plots
of the response density  $\delta n(\omega,r)$  (\ref{eq-nw}) in Figure~\ref{fig-cnts},
which have been calculated for the specific plasmon resonances.
The plasmon velocity increases to $2.54$ times the Fermi velocity
(here at $10^6$ m/s) for the 40 unit cells  (3,3)-CNT, and it is expected to level off
if we keep increasing the length of the CNT.

%%%%%%%%%%%%%%%%%%%%%%%%%%%%%%%%%%%%%%%%%%%%%%%%%%%%%%%%%%%%%%%%%%%%%%%%%%%%%%
%%%%%%%%%%%%%%%%%%%%%%%%%%%%%%%%%%%%%%%%%%%%%%%%%%%%%%%%%%%%%%%%%%%%%%%%%%%%%%%

\section{Conclusion}

Modern first-principle calculations aim at bringing
 computational activities up to the level where they can significantly impact innovations in
electronic nanomaterials and devices research.
Nanostructures with many atoms and electrons can only be treated by addressing the efficiency and scalability of algorithms
on modern computing platforms with multiple hierarchical levels of parallelism.
These goals can be achieved using an efficient
 modeling framework that can perform real-space DFT ground-state calculations, and real-time TDDFT excited-state calculations.
 The latter can be used to study  various relevant quantum many-body effects (such as
plasmonic effects) by performing
electronic spectroscopy.
Spectroscopic techniques are among the most fundamental probes of matter: incoming radiation
 perturbs the sample and the response to this perturbation is measured.
The system is inherently excited in this process and hence, a calculation of ground-state properties
is insufficient to interpret the response of the system.
 TDDFT has had considerable success modeling the interaction of
electromagnetic fields with matter, and obtaining spectroscopic information with absorption and emission spectra.

The paper discussed the NESSIE software which has been fundamentally designed to take advantage of parallel optimization
at various levels of the entire modeling process. NESSIE benefits from the linear
scaling capabilities of real-space mesh techniques and domain decomposition methods
to perform all-electron (full core potential) calculations.
The modeling approach is tailored to optimally take advantage of the full capability of the
state-of-the-art FEAST eigensolver that can achieve significant parallel scalability
on modern HPC architectures. With the  success in meeting these challenges, 
NESSIE is currently able to extend the first-principle simulations to very large atomistic structures
(i.e. many thousands electrons at the level of all-electron/DFT/TDDFT/ALDA theory).
The modeling framework opens new perspectives for addressing the numerical challenges in
TDDFT excited-state calculations to operate the full range of
electronic spectroscopy, and study the nanoscopic many-body effects in arbitrary complex molecules and
  finite-size large-scale nanostructures.
It is expected that the NESSIE software and associated numerical components can
become a new valuable new tool for the scientific community, to investigate
the fundamental electronic properties of numerous nanostructured materials.

\section*{Acknowledgment}

This work was also supported
by the National Science Foundation,
under grants \#CCF-1510010 and SI2-SSE\#1739423.
The CNT calculations used the Extreme Science and Engineering
Discovery Environment (XSEDE), which is supported by National Science Foundation grant number ACI-1548562.

\ifCLASSOPTIONcaptionsoff
  \newpage
\fi



%\bibliographystyle{IEEEtran}


\begin{IEEEbiographynophoto}{James Kestyn}
James Kestyn received his PhD in Electrical Engineering at the University of Massachusetts, Amherst in 2018. 
His graduate studies focused on electronic structure calculations, where he helped to develop the multi-level parallel approach used in NESSIE, and on numerical linear algebra, developing the non-Hermitian extension of the FEAST eigenvalue solver.
He has held internships at Intel and Samsung and has since been working on computational electromagnetics at Stellar Science Ltd Co.
\end{IEEEbiographynophoto}

\begin{IEEEbiographynophoto}%[{\includegraphics[width=1in,height=1.25in,clip,keepaspectratio]{polizzi}}]
	{Eric Polizzi}
Eric Polizzi is Professor in the Department of Electrical
and Computer Engineering and the Department of Mathematics and Statistics at the University of Massachusetts,
Amherst. He received an education in theoretical and computational physics, and a
PhD in Applied Mathematics (2001) from the University of
Toulouse, France. Prior to joining UMass in 2005, he served
as a postdoctoral research associate in Electrical
Engineering (2002-2003) and as a senior research scientist in
Computer Sciences (2003-2005), both at
Purdue University. Prof. Polizzi is conducting interdisciplinary research
activities at the intersection between advanced
mathematical techniques, parallel numerical algorithms, and
computational nanosciences. These activities can be broadly divided into
three projects: the  DFT/TDDFT NESSIE code, the
parallel linear system solver SPIKE, and the eigenvalue solver FEAST.
\end{IEEEbiographynophoto}

%\newpage


%\vfill


\end{document}